 \documentclass[final,authoryear,3p,times,fleqn]{elsarticle}
\usepackage{amssymb}
\usepackage{amsmath}
\usepackage{amsthm}
\usepackage{mathpazo}
\usepackage[ruled]{algorithm2e}
\usepackage{derivative}

\usepackage{graphicx}
\usepackage{float} 
\usepackage{pxfonts,times}
\usepackage{siunitx}

\usepackage{bm}
\usepackage{color}
\usepackage{comment}
\usepackage{natbib}
\usepackage[ruled]{algorithm2e}

\usepackage[
	colorlinks=true, 
	citecolor=blue, 
	urlcolor=blue]{hyperref}
\usepackage{listliketab}
\usepackage{booktabs}
\usepackage{tabularx}
\newcolumntype{C}{>{\centering\arraybackslash}X}
\newcolumntype{L}{>{\raggedright\arraybackslash}X}
\newcolumntype{R}{>{\raggedleft\arraybackslash}X}
\usepackage{longtable}

\makeatletter
\def\th@plain{\upshape}
\makeatother

\makeatletter
	
	\@addtoreset{equation}{section}
\makeatother

\journal{}

\begin{document}

\begin{frontmatter}



\title{Global path preference and local response: A reward decomposition approach \\for network path choice analysis in the presence of locally perceived attributes}
%A Markovian network\\ path choice model with decomposed reward functions
%A reward decomposition approach\\ for link-based network path choice modeling
%A reward decomposition approach for network path choice analysis
%Network route choice modeling with decomposed Markov utilities
%Learning network path choice behavior with decomposed Markov rewards
%A reward decomposition approach to network path choice modeling
%A recursive route choice model with decomposed Markov rewards

%\tnoteref{label1}
%\tnotetext[label1]{Acknowledgements. The authors express their gratitude to the anonymous referees for their careful reading of the manuscript and useful suggestions. The first author was supported by Kajima Foundation. The first and fourth authors were supported by JSPS KAKENHI Grants No.15H04053 and No.16H02368.}


\author{Yuki Oyama} %\corref{cor1}
\address{Department of Civil Engineering, Shibaura Institute of Technology, Tokyo, Japan}
\ead{oyama@shibaura-it.ac.jp}
\cortext[cor1]{Corresponding author}

\begin{abstract}
This study performs an attribute-level analysis of the global and local path preferences of network travelers. To this end, a reward decomposition approach is proposed and integrated into a link-based recursive (Markovian) path choice model. 
The approach decomposes the instantaneous reward function associated with each state-action pair into the global utility, a function of attributes globally perceived from anywhere in the network, and the local utility, a function of attributes that are only locally perceived from the current state. Only the global utility then enters the value function of each state, representing the future expected utility toward the destination. 
This global-local path choice model with decomposed reward functions allows us to analyze to what extent and which attributes affect the global and local path choices of agents. Moreover, unlike most adaptive path choice models, the proposed model can be estimated based on revealed path observations (without the information of plans) and as efficiently as deterministic recursive path choice models. The model was applied to the real pedestrian path choice observations in an urban street network where the green view index was extracted as a visual street quality from Google Street View images. The result revealed that pedestrians locally perceive and react to the visual street quality, rather than they have the pre-trip global perception on it. Furthermore, the simulation results using the estimated models suggested the importance of location selection of interventions when policy-related attributes are only locally perceived by travelers.
\end{abstract}


\end{frontmatter}

% \linenumbers

\section{Introduction}
\begin{figure}[t] %tb
	\begin{center}
		\includegraphics[width=14cm]{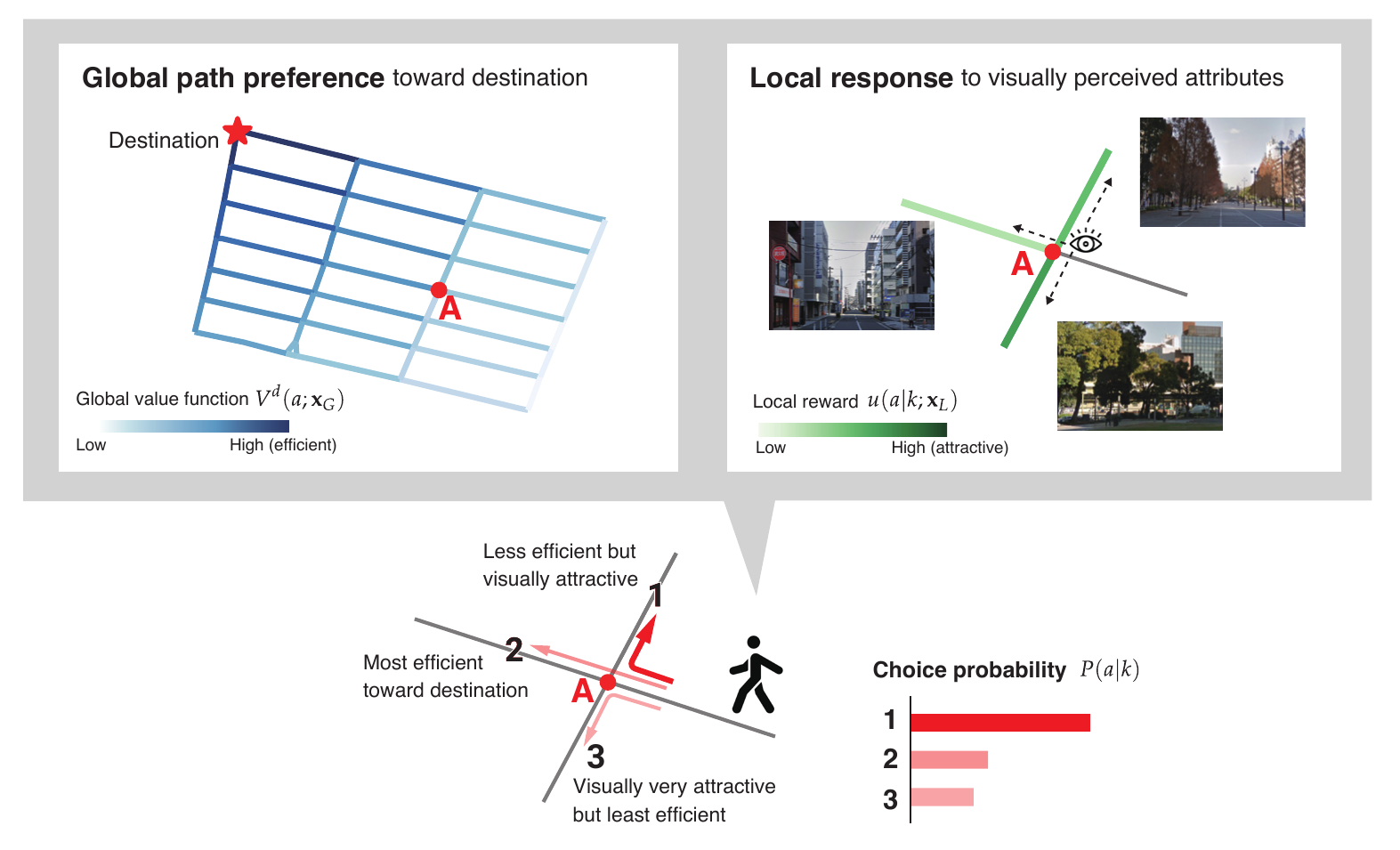}
		\caption{Conceptual diagram of the global path preference based on pre-trip information and local response to visually perceived attributes, captured by the global-local path choice model (the pictures are borrowed from Google Street View images). In this example, a pedestrian responds to visual street qualities en route and is likely to choose link 1, which is globally less efficient to reach the destination but visually more attractive than link 2.}
		\label{fig:diagram} 
	\end{center}
\end{figure}

% \subsection{Background and motivation}
A network path choice model predicts which path an agent travels between an origin-destination (OD) pair on a network represented by a directed graph. While a traveler chooses a path to the destination based on the \textit{pre-trip} global perception of network attributes, obtained from past experience or external information, s/he can also visually perceive local network conditions \textit{en route} and adjust the path at intermediate nodes.
% observe the local environment of a network while traveling and myopically adjust their paths. 
This myopic response of travelers to locally perceived network attributes can be observed for various types of networks when travelers face \textit{unexpected events}; to name a few, (1) drivers perceive actual conditions of road segments ahead (e.g., road closure, incidents, and disruption) and change their plans \citep{como2013stability, gao2010adaptive}; (2) cyclists adjust their paths according to road surface conditions or traffic lights \citep{stinson2003commuter}; (3) pedestrians are attracted by the visual quality of a street and myopically choose to walk on it \citep{Oyama2012, natapov2016visibility}. %; (4) during a disaster evacuee locally perceive network disruption and are forced to change their plans \citep{gao2010adaptive} 
%travelers locally adjust their paths while trying to efficiently reach their destination (e.g., expected shortest paths). In other words, 
As such, travelers' path choice behavior is based on two routing mechanisms: \textbf{global path preferences} for and \textbf{local responses} to perceived network attributes (Figure \ref{fig:diagram}). 
Analyzing local responses to the network environment revealed en route as well as global path preferences is essential to well describe realistic behavior and network traffic dynamics. 
% Although route choice modeling literature has mainly focused on the analysis of global path preferences by modeling \textit{pre-trip} choices of paths, capturing local responses to the network environment revealed during the travel is important to well describe realistic behavior and network traffic dynamics. % \citep{como2013stability}
% Network traffic dynamics are influenced by the local responses as well as global path choices \citep{como2013stability}.
% As such, modeling different routing mechanisms of \textbf{global path preferences} and \textbf{local responses} is essential in analyzing realistic path choice behavior. 

Modeling adaptive path choice behavior has been extensively studied in the context of real-time travel information provision, where drivers revise their path choices with new information provided en route \citep[e.g.,][]{abdel1997using, mahmassani1999dynamics, peeta2005hybrid, de2020route}. % regarding travel times
However, travelers' responses to local environments have yet to be sufficiently analyzed in the other applications, such as situations where \textit{slow mode} travelers (e.g., pedestrians, cyclists, or micromobility users) locally react to visually perceived network attributes, on which this paper particularly focuses in the case study. %despite its importance, 
%this type of behavior has yet to be sufficiently analyzed in other applications than real-time travel information
% This study aims at modeling the local path choice behavior, with a particular focus on situations that \textit{slow mode} travelers (e.g., pedestrians or cyclists) locally respond to visually perceived network attributes. These types of behavior have yet to be sufficiently understood, while the literature on en route path choice modeling has mainly focused on the real-time information provision regarding travel times to drivers \citep[e.g.,][]{abdel1997using, mahmassani1999dynamics, ding2019latent, de2020route}. 
% The real-time provision of travel information is not the main focus of this study although it has been the main motivation behind the literature of en route path choice modeling \citep[e.g.,][]{abdel1997using, mahmassani1999dynamics, de2019modelling, de2020route}. 
Compared to car drivers, slow-mode travelers are potentially more sensitive to the visual environment en route and locally perceive more network attributes (e.g., streetscape or unexpected road surface conditions). On the other hand, they can still globally consider some attributes (e.g., path length) for choosing a path efficiently leading them to the destination. Therefore, an attribute-level analysis of global and local preferences is necessary for understanding to what extent and which attributes locally affect the path choices of travelers and designing related policies.

In this paper, we present a novel network path choice model to analyze the traveler's global preferences for and local responses to various network attributes from revealed preference (RP) data. Figure \ref{fig:diagram} provides the conceptual diagram of the model.
We consider a situation where a traveler visually perceives some attribute of links outgoing from the current link and responds to the locally updated condition.
%takes an action that locally maximizes his/her utility. 
%consider a specialized case where
%responds to the updated condition
For this purpose, we propose a \textbf{reward decomposition} approach incorporated into the formulation of a link-based recursive logit (RL) model that describes a network path choice as sequential link choices toward the destination in a Markovian fashion \citep{Fosgerau2013RL, Mai2015NRL, Oyama2017GRL}. 
%based on the formulation of a link-based recursive path choice model \citep{Fosgerau2013RL}, 
The approach decomposes the Markov reward (instantaneous utility) function associated with each link (state-action) pair into \textbf{global utility} that is globally perceived from anywhere in the network and \textbf{local utility} that is only locally perceived when the traveler arrives at the intersection connecting the links.
The value function of each state, i.e., the future expected maximum utility to the destination, results in the function of only the global utility and represents the path preferences of a traveler. 
Moreover, unlike most of the previous adaptive or local path choice models, our model can be estimated only with revealed path observations. Thus, the proposed decomposition approach allows us to empirically analyze the global and local preferences for each network attribute. 

As an application of the proposed model, this study shows a case study in a real pedestrian network and with walking path observations from GPS data. The green view index (GVI) value of streets, extracted from Google Street View images by a semantic segmentation algorithm, is introduced as a locally perceived network attribute with the expectation that the visual street quality affects pedestrians' decisions of walking streets en route. The model estimation results reveal that GVI is locally perceived and positively affects pedestrians' sequential path choices. Furthermore, the simulation results using the estimated models suggest the importance of location selection of interventions when the attributes of interest are only locally perceived by travelers.

The remainder of the paper is structured as follows. Section \ref{sec:review} reviews the related path choice models. Section \ref{sec:model} introduces the modeling framework with an illustrative example. Section \ref{sec:estimation} discusses the model estimation. Section \ref{sec:result} presents several numerical results based on both synthetic data and real observations. Section \ref{sec:conclusion} concludes and discusses future directions. \ref{app:gradient} provides the detailed deviation of the gradients of the likelihood function, and \ref{app:discount} shows the results of discounted models in the real case study.

%%%%%%%%%%%%%%%%%%%%%%%%%%%%%%%%%%%%%
%%%Figure of Diagram
%%%%%%%%%%%%%%%%%%%%%%%%%%%%%%%%%%%%%
%%%%%%%%%%%%%%%%%%%%%%%%%%%%%%%%%%%%%

\section{Literature review}\label{sec:review}
Most network path choice models in the literature describe the traveler's pre-trip decision on a path between an OD pair, analyzing global path preferences \citep[see e.g., the reviews by][for the overviews]{prato2009route, oyama2022markovian, duncan2020path}. Such \textit{fixed} path choice models do not account for the traveler's local response to the updated perception of network attributes. This literature review focuses on three types of path choice models that describe en-route or myopic decisions, categorized into \textit{adaptive}, \textit{link-based}, and \textit{local} path choice models.

\subsection{Adaptive path choice models}
%Travelers’ route choice behavior in an uncertain network with real-time information will conceivably be different from that in a deterministic network. With real-time information provided en route, travelers could make route choice decisions at intermediate nodes based on the current situation in order to avoid delay downstream. An adaptive path model assumes that a traveler is reactive and route choice is a series of path choices at each node.
Adaptive path choice models have been studied mainly for real-time information provision to drivers, describing the driver's path choice decisions at intermediate nodes toward the destination given real-time information provided en route. 
Models in this category are often based on the \textit{plan-action} modeling framework \citep{choudhury2010dynamic}: a traveler is assumed to make a \textit{plan} based on the pre-trip information and take an \textit{action} reacting to the traffic conditions revealed en route. %, travel along the chosen path, perceive traffic conditions revealed en route,
Many studies modeled this type of behavior as route-switching, where plan represents a path that a traveler decides before the trip or is currently traveling, and action is the switching to another path alternative at each intermediate decision node \citep[e.g.,][]{polydoropoulou1996modeling, abdel1997using, mahmassani1999dynamics}. These models have been implemented in dynamic traffic simulators such as DYNASMART \citep{mahmassani2001dynamic} and DynaMIT \citep{ben2002real}.
% The route-switching model is one of the mainstreams in this context \citep{abdel1997using, mahmassani1999dynamics}. A route-switching model predicts if a traveler obtaining new information en route switches from the planned path to an alternative. 

Another modeling approach of adaptive path choice behavior is a routing policy choice model, where a routing policy is the traveler's plan that maps from all possible states to actions on which link to take next \citep{gao2008adaptive, gao2010adaptive, razo2013rank}. Under the uncertainty of travel times, travelers are assumed to choose a routing policy before the trip and adapt to traffic conditions revealed en route by executing the plan. Thus, routing policy choice models can analyze the strategic behavior of travelers, while route-switching models describe myopic and successive switchings of path choices.
% While route-switching models describe myopic and successive switchings of path choices, some studies modeled the strategic behavior in the traveler's adaptive path choice as routing policy choice models, where a routing policy is the traveler's plan that maps from all possible states to actions on which link to take next \citep{gao2008adaptive, gao2010adaptive}. 
% Under the uncertainty of travel times, travelers are assumed to choose a routing policy before the trip and adapt to traffic conditions revealed en route by executing the plan.

\subsection{Link-based recursive path choice models}
Link-based recursive path choice (RL) models describe the traveler's path choice behavior as sequential link choices under a Markov decision process (MDP) framework \citep{Fosgerau2013RL, Mai2015NRL}. The link-based formulation based on a dynamic discrete choice model \citep{Rust1987} allows us to implicitly consider the unrestricted path set, thereby providing a computationally efficient way of modeling and a consistent estimator. Although the RL model originally describes static and global path choice behavior, some recent studies presented its extensions to model myopic or adaptive path choice behavior, focusing on the sequential decision-making structure of MDP.
\cite{Oyama2017GRL} presented a discounted RL model to analyze myopic decision-making processes during a disaster where drivers do not have sufficient information about network conditions. \cite{de2020route} analyzed the effect of travel information provided en route by modeling an RL model in a deterministic time-space network. 
\cite{mai2021route} formulated a link-based routing policy choice model as an RL model in a stochastic network, which addressed the choice set generation problem of path-based routing policy choice models \citep{ding2019latent}. Although \cite{mai2021route} proposed a solution algorithm for tractable computation, the computational effort required for the stochastic model is still much higher than that for deterministic RL models.

As highlighted in \cite{zimmermann2020tutorial}, the RL path choice models are mathematically and closely related to the maximum entropy inverse reinforcement learning (IRL) model \citep{ziebart2008maximum}. Recently, \cite{zhao2023deep} presented a deep IRL model for path choice modeling, where context-dependent rewards were introduced to capture global trip contexts such as trip purpose, socio-demographic characteristics, and destination, not local contexts.

\subsection{Local path choice models}
Local path choice models describe the myopic decisions of travelers who perceive the local environment during travel to the destination. 
These myopic path choice behavior are often studied in the context of pedestrian modeling \citep{antonini2006discrete, robin2009specification, Oyama2012, Oyama2018ped}
% , including path choices in urban street networks \citep{tsukaguchi2001modelling, Oyama2012, Oyama2018ped} or microscopic behavioral simulations \citep{antonini2006discrete, robin2009specification}, 
where pedestrians are assumed to choose the next link/step to move based on locally perceived spatial attributes, while global routing preferences are not explicitly considered but are simplified to the orientation to the destination.
% as well as the orientation to the destination.
% simulation models, where they choose the next step (including speed or direction) while directing to the destination \citep{antonini2006discrete, robin2009specification}. 

In contrast, \cite{hoogendoorn2015continuum} proposed a pedestrian traffic flow theory by modeling both global and local path choices: the global path choice represents the pre-trip decision based on the expected flow conditions, and the local path choice reflects the adaptation to local conditions around a pedestrian. This approach mainly addressed the challenges of previous global path choice models \citep{hoogendoorn2004pedestrian}, such as expensive computational effort and unrealistic behavioral assumptions that pedestrians globally forecast their movements with each other. However, since their framework is designed for pedestrian traffic simulation on a continuous space, it does not describe discrete path choices in a network or is not empirically estimated with real trajectories.

% local path choices, within the framework of optimal 

\subsection{Positioning and contributions of the study}
In the literature, path choice models with travelers' responses to local contexts are mainly designed for real-time travel information provision. Therefore, most models only consider traffic conditions and resultant travel times for network attributes. Moreover, a plan (strategy) of a traveler is generally latent and thus cannot be directly observed, and only a few studies of adaptive path choice models presented the empirical estimation using RP data \citep{ding2019latent, mai2021route}, while many studies relied on laboratory experiments or stated-preference (SP) data \citep[e.g.,][]{abdel1997using, mahmassani1999dynamics, razo2013rank}.

This study proposes a link-based recursive path choice model with the instantaneous reward decomposed into global and local utilities, which we name a \textit{global-local path choice model}. 
This framework has a number of advantages: (1) it can capture both the global and local preferences of travelers for different network attributes; (2) it can be estimated with RP data of realized path observations; (3) the required computational effort is as less as the deterministic RL models \citep{Fosgerau2013RL, Mai2015NRL, Oyama2017GRL}.
% As such, it allows us to empirically analyze to what extent and which attributes affect the local responses of travelers in a network. 
% These points also make important differences to the pedestrian traffic simulation model of \cite{hoogendoorn2015continuum}, although our model is conceptually similar to their framework.
% Although it is conceptually similar to the framework of \citep{hoogendoorn2015continuum}, our model is a discrete choice on a network and can be empirically estimated.
% which can analyze both the global and local preferences of travelers for different network attributes. It can also be estimated with RP data of realized path observations. 
While our objective is to empirically analyze to what extent and which attributes affect travelers' local responses in a network, the third item above implies that our model can also be viewed as a reduced version of stochastic MDP models \citep[e.g.,][]{mai2021route}, which are costly and often have difficulty in defining the distributions of stochastic attributes (i.e., state transitions).
In addition, we show an application of the model to pedestrian path choices in an urban street network and reveal that they locally respond to visual street quality while having global path preferences.
% and analyze their local responses to visually perceived network attributes. 
The pedestrian path choice analysis in an urban network has been increasing attention in the context of walkability, and this study gives new findings on their local responses that previous path choice analyses could not capture \citep{erath2015modelling, basu2022street, isenschmid2022zurich, oyama2023capturing}. As such, our study opens up new application fields of global-local path choice modeling.
% and is a new application field of local path choice modeling. Therefore, our study opens up various future research directions. 

\section{Global-local path choice model}\label{sec:model}
This section presents the global-local path choice model, a link-based recursive path choice model integrated with a reward decomposition approach. 
We assume that a traveler on a link visually perceives some attributes of the outgoing links from the current link\footnote{This is a similar setting to \cite{gao2010adaptive} and \cite{como2013stability}, and this study deals with various attributes, not limited to travel times affected by traffic conditions.} and responds to the updated condition so that his/her utility is locally maximized. 

% The objective of a traveler is to maximize the reward obtained along the travel to the destination. Therefore, a traveler in state $a_j$ is assumed to choose a path $\{a_j, \ldots, a_J\}$ that maximizes the accumulated reward:
% \begin{equation}
%     \label{eq:path_reward}
%     u(\{a_j, \ldots, a_J\}) = \sum_{t=j}^{J-1} \gamma^{t-j} u(a_{t+1}|a_t)
%     %u(\{a_j, \ldots, a_J\}) = u_L(a_{j+1}|a_j) \sum_{t=j}^{J-1} \gamma^{t-j} u_G(a_{t+1}|a_t)
% \end{equation}
% where $\gamma$ is the discount factor of future rewards.

\subsection{Markov decision process and reward decomposition}
Consider a directed graph $G = (N, L)$, where $N$ is the set of nodes and $L$ is the set of links. This study describes the path choice behavior of a traveler as a sequential decision process in the network, based on an MDP \citep{ziebart2008maximum, Fosgerau2013RL}. A \textit{state} of the MDP is defined as a link $k \in L$ of the network, and \textit{action} is the choice of a subsequent link $a \in A(k)$ to move on, where $A(k) \subset L$ is the action set available to a traveler in state $k$. In other words, a traveler directly chooses the next state by taking an action, thus the MDP is deterministic. An action $a \in A(k)$ given the current state $k$ is associated with a perceived \textit{reward} (utility) $u(a|k)$ for a traveler, which represents the traveler's routing preferences. 
%To describe the sequential decision-making process of a traveler in the network, this study formulates the path choice model based on a Markov decision process (MDP) \citep{ziebart2008maximum, Fosgerau2013RL}.  

The core idea of this study is the decomposition of the reward function $u(a|k)$ to simultaneously describe both global path preferences for and local responses to network attributes. Specifically, we define $u(a|k)$ as the sum of \textit{global} utility $u_G(a|k)$ and \textit{local} utility $u_L(a|k)$:
\begin{equation}
    \label{eq:decompose_u}
    u(a|k) = u_G(a|k) + u_L(a|k).
\end{equation}
The global utility $u_G(a|k)$ is perceived by a traveler in any state $s \in L$ in the network, thus describing the global preferences for a path to the destination. In contrast, the local utility $u_L(a|k)$ of link $a \in A(k)$ is only perceived by a traveler in state $k$ and is assumed to be zero for the other state $s \in L \setminus \{k\}$. In other words, a traveler considers the utility $u_L(a|k)$ only for the decision of action $a \in A(k)$ after arriving at link $k$. 

Each of the global and local utilities is further decomposed into its deterministic and error components:
\begin{subequations}
    \begin{align}\centering
        u_G(a|k) &= v_G(a|k) + \epsilon_G(a|k), \label{eq:ug}\\
        u_L(a|k) &= v_L(a|k) + \epsilon_L(a|k), \label{eq:ul}
    \end{align}
\end{subequations}
where $v_G(a|k) = v(\mathbold{x}_{G,a|k}, \beta_G)$ is a function of the \textit{globally perceived attributes} $\mathbold{x}_{G,a|k}$, and $v_L(a|k) = v(\mathbold{x}_{L,a|k}, \beta_L)$ is a function of the \textit{locally perceived attributes} $\mathbold{x}_{L,a|k}$. The parameter vectors $\beta_G$ and $\beta_L$ are the weights of attributes to be learned from data.
We assume the extreme value (EV) distribution with scale $\mu_G > 0$ for the error component of the global utility $\epsilon_G(a|k)$, and another EV distribution with scale $\mu > 0$ for the sum of error components $\epsilon(a|k) = \epsilon_G(a|k) + \epsilon_L(a|k)$.
%$ \sim {\rm EV}(0, \mu_G)$, $ \sim {\rm Gumbel}(0, \mu)$

\subsection{Global value function}\label{sec:valuefunc} % and local path choice model
At the sink node of link $k$, perceiving the local utilities $u_L(a|k)$ of the outgoing links $a \in A(k)$, a traveler takes an action so that the sum of the global value function $V^d(a)$ plus the local utility $u_L(a|k)$ is maximized. The global value function $V^d(k)$ of state $k$ represents the expected maximum utility of possible paths from link $k$ toward the destination $d$:
\begin{equation}
    \label{eq:EMU}
    V^d(k) \equiv \mathbb{E}\left[\max_{r \in R_{kd}} \{u_G(r)\} \right],
\end{equation}
where $R_{kd}$ is the set of all feasible paths departing from $k$ and terminating at $d$ in the network, and the (discounted) utility $u_G(r)$ of path $r$ is defined by
\begin{equation}
    \label{eq:path_util}
    u_G(r = \{a_1, \ldots, a_J\}) = \sum_{t=1}^{J-1} \gamma^{t-1} u_G(a_{t+1}|a_t),    
\end{equation}
with $a_1 = k$ and $a_J = d$, and $\gamma \in (0,1]$ is the discount factor of future utilities. 
The global value function (\ref{eq:EMU}) can be recursively formulated via Bellman's equation:
%A traveler in state $k$ is assumed to locally optimize her/his choice by maximizing the sum of the local utility $u_L(a|k)$ and the global value function $V^d(a)$ describing the expected path utility toward destination $d$. The global value function $V^d(k)$ of state $k$ is recursively formulated via Bellman equation:
%choose the action $a$ that maximizes the sum of instantaneous reward $u(a|k)$ and the expected downstream reward $V^d(a)$ to the destination (absorbing state) $d$. The expected utility $V^d(a)$ is the \textit{global} value function of state $k$ that is formulated via Bellman equation:
\begin{equation}
    \label{eq:bellman}
    V^d(k) \equiv \mathbb{E}\left[\max_{a \in A(k)} \{u_G(a|k) + \gamma V^d(a)\} \right],
    % + \epsilon_G(a|k)
\end{equation}
and $V^d(d) = 0$.
This value function formulation is the same as that of the RL models \citep{Fosgerau2013RL, Oyama2017GRL}, which are global path choice models formulated based on the dynamic discrete choice modeling framework \citep{Rust1987}. %for global path choice modeling
Nevertheless, the key difference is that we decompose the reward function and consider that only a part of the reward affects the global path choice. As a result, the value function (\ref{eq:bellman}) is evaluated based only on the global utility $u_G(a|k)$ and does not depend on local utility $u_L(a|k)$; i.e., $V^d(k) = V^d(\mathbold{v}_G(\cdot|k), \mu_G)$.
%As shown, $V^d(k) = V^d(\mathbold{v}_G(\cdot|k), \mu_G)$ is a function of global utility $\mathbold{v}_G(\cdot|k)$ and does not depend on local utility $\mathbold{v}_L(\cdot|k)$.
%We define the global value function $V^d(k)$ describing the expected maximum reward associated with the travel from state $k$ toward destination $d$:

With the distributional assumption on $\epsilon_G$, Eq.(\ref{eq:bellman}) further reduces to the following logsum function:
\begin{equation}
    \label{eq:logsum}
    V^d(k) = \frac{1}{\mu_G} \ln \sum_{a \in A(k)} e^{\mu_G \{v_G(a|k) + \gamma V^d(a)\}},
\end{equation}
and equivalently,
\begin{equation}
    \label{eq:elogsum}
    e^{\mu_G V^d(k)} = \sum_{a \in A(k)} e^{\mu_G \{v_G(a|k) + \gamma V^d(a)\}}.
\end{equation}
Because the network path choice problem considers the destination $d$ (i.e., the absorbing state) of a traveler to be given, the MDP is episodic and the discount factor $\gamma$ is often assumed to be one \citep{zhao2023deep}. In other words, path choice MDPs usually deal with an undiscounted case \citep{Akamatsu1996MCA, Fosgerau2013RL, mai2022undiscounted}, on which this study also mainly focuses. 
In such cases, the value functions can be efficiently solved through a system of linear equations:
%which is a system of linear equations that can be efficiently solved \citep{Fosgerau2013RL}.
\begin{equation}
    \label{eq:linearsystem}
    \mathbold{z}^d = \mathbf{M}\mathbold{z}^d + \mathbold{b}^d \Leftrightarrow \mathbold{z}^d = (\mathbf{I} - \mathbf{M})^{-1} \mathbold{b}^d
\end{equation}
where $z^d_k = e^{\mu_G V^d(k)}$; $M_{ka} = \delta(a|k) e^{\mu_G v_G(k|a)}$; $\delta(a|k)$ is the state-action incidence taking one if $a \in A(k)$ and zero otherwise; and $b^d_k$ equals one if $k = d$ and zero otherwise. 
Note that for a discounted case, i.e., when $\gamma < 1$, the system becomes non-linear and the value iteration can be applied to solve the value function \citep{Oyama2017GRL}. %, but less efficient compared to the linear system

\subsection{Locally optimal behavior and path choice probability}
The global value function represents the expected and accumulated rewards toward destination $d$, describing the path preferences of a traveler.
The local utility affects this global path choice and leads to locally optimal behavior in each state, resulting in the adaption of path choice to the local conditions of the environment (Figure \ref{fig:diagram}).
Given the distributional assumption on $\epsilon$, the probability of a traveler in state $k$ to take an action $a$ is
\begin{equation}
    \label{eq:prob}
    p^d(a|k) = \frac{
        e^{\mu \{v(a|k) + \gamma V^d(a)\}}
    }{
        \sum_{a' \in A(k)} e^{\mu \{v(a'|k) + \gamma V^d(a')\}}
    },
\end{equation}
reflecting the traveler's decision of maximizing the sum of the local utility and global value function.
To make the difference to the global path choice models \citep{Fosgerau2013RL, Oyama2017GRL} clear, we can expand the right-hand side of (\ref{eq:prob}) as
\begin{align}\centering
    \label{eq:prob_ex}
    p^d(a|k)
    = \frac{
        e^{\mu \{v_G(a|k) + v_L(a|k) + \gamma V^d(a)\}}
    }{
        \sum_{a' \in A(k)} e^{\mu \{v_G(a'|k) + v_L(a'|k) + \gamma V^d(a')\}}
    } 
    = \frac{
        e^{\mu \{v_G(a|k) + v_L(a|k)\}} (z^d_a)^{\frac{\gamma \mu}{\mu_G}}
    }{
        \sum_{a' \in A(k)} e^{\mu \{v_G(a'|k) + v_L(a'|k)\}} (z^d_{a'})^{\frac{\gamma \mu}{\mu_G}}
    }. 
    % &= \frac{
    %     M_{l,ka} (M_{g,ka})^{\frac{\mu_G}{\mu}} (z^d_a)^{\frac{\mu_G}{\mu}}
    % }{
    %     \sum_{a' \in A(k)} M_{l,ka'} (M_{g,ka'})^{\frac{\mu_G}{\mu}} (z^d_{a'})^{\frac{\mu_G}{\mu}}
    % }
\end{align}
This model corresponds to the global path choice model $P^d_G(a|k)$ as a special case when $v_L(a|k) = 0$ and $\mu = \mu_G$, i.e., when all the network attributes are globally perceived by travelers:
\begin{align}\centering
    \label{eq:Pg}
    P^d_G(a|k)
    = \frac{
        e^{\mu_G \{v_G(a|k) + \gamma V^d(a)\}}
    }{
        \sum_{a' \in A(k)} e^{\mu_G \{v_G(a'|k) + \gamma V^d(a')\}}
    },
\end{align}
which further reduces to $P^d_G(a|k) = M_{ka}z^d_a/z^d_k$ when $\gamma = 1$.

Because the path choice from origin to destination is the outcome resulting from the local choice process, the probability of path $r = \{a_1, \ldots, a_J\}$ with $a_J = d$ is the product of action choice probabilities
\begin{equation}
    \label{eq:path_prob}
    P_r = \prod^{J-1}_{j=1} p^d(a_{j+1}|a_j).
\end{equation}

\subsection{Illustrative example}
\begin{figure}[t] %tb
	\begin{center}
		\includegraphics[scale=1.0]{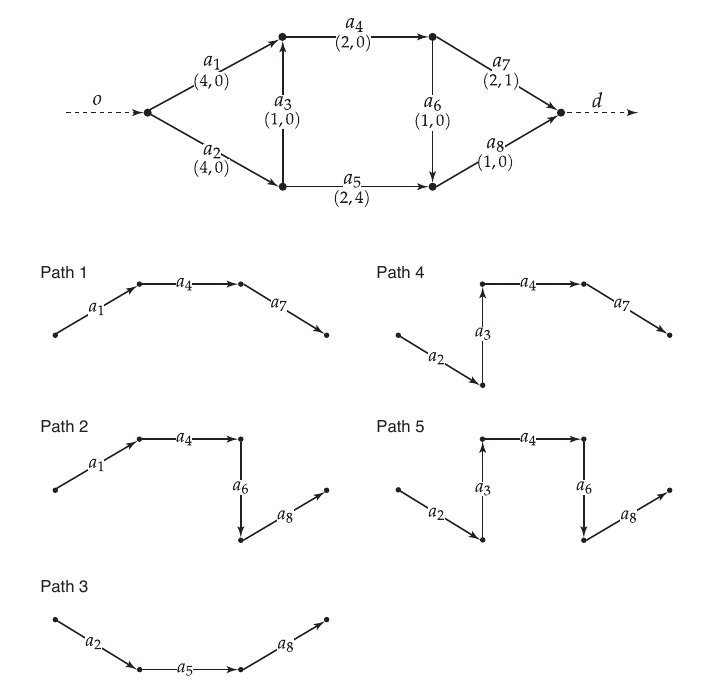}
		\caption{An example network and path alternatives. The numbers in the parentheses (below the link number) on each link indicate the link attribute vector $\mathbold{x}_{a} = (x_{1,a}, x_{2,a})$.}
		\label{fig:laddernet} 
	\end{center}
\end{figure}

To show how to specify the global-local path choice model and what the model can describe, this subsection presents an illustrative example using the network of Figure \ref{fig:laddernet}. Five path alternatives are available to travelers between origin $o$ to destination $d$, and their probabilities are evaluated using the proposed global-local path choice model (\ref{eq:prob}). 
We consider two attributes $\mathbold{x}_a = (x_{1,a}, x_{2,a})$ for each link and define three test cases based on how these attributes are perceived by travelers. Table \ref{tb:cases} shows the three cases: in case 1, we only consider $x_{1}$ that is globally perceived; in case 2, both $x_{1}$ and $x_{2}$ are considered and perceived globally; and in case 3, $x_{2}$ is perceived only locally while $x_{1}$ is globally perceived. 

%%%%%%%%%%%%%%%%%%%%%%%%%%%%%%%%%%%%%
%%%Figure of Network for Illustrative Example
%%%%%%%%%%%%%%%%%%%%%%%%%%%%%%%%%%%%%
%%%%%%%%%%%%%%%%%%%%%%%%%%%%%%%%%%%%%

For the interpretation simplicity, let us say that $x_{1}$ is the expected link travel time, and $x_{2}$ is the additional link travel time caused by an unexpected event (e.g., traffic jam, road repairing, incident). In this setting, case 2 represents a situation where travelers have obtained the information about the additional travel time in advance (say, through a mobile app), while in case 3 travelers do not have access to the information or know until arriving at the link.

Table \ref{tb:probs} reports the path probabilities for the three cases, where we set the scales of the error term distributions to $\mu = \mu_G = 1$ and the discount factor to $\gamma = 1$. In case 1 where additional travel times were not considered, path 3 had the highest probability ($P_3 = 0.498$), followed by paths 1 and 2 ($P_1 = P_2 = 0.183$), then paths 4 and 5 ($P_4 = P_5 = 0.067$).
In case 2 where an unexpected event occurred and travelers were informed of the incurred additional travel times before the departure, most of the travelers gave up choosing path 3 and changed their plans. As a result, path 2 avoiding the links $a_5$ and $a_7$ whose travel times increased got the highest choice probability ($P_2 = 0.521$), followed by paths 1 and 5 ($P_1 = P_5 = 0.192$), path 4 ($P_4 = 0.070$) and path 3 ($0.026$). Note that the results in cases 1 and 2 are consistent with those of the non-adaptive and path-based multinomial logit (MNL) model.

In contrast, in case 3, travelers could not obtain information about the additional travel times before the trip, and they had to adjust their paths locally. Because they globally perceived only $x_1$ when they departed from the origin $o$, they originally planned their paths according to the same path probabilities as case 1, resulting in more than half of travelers taking an action to travel link $a_2$ that is the first elemental link of path 3 (and paths 4 and 5). However, at the sink node of link $a_2$, the travelers locally perceived the additional time of link $a_5$ and switched their paths, moving to link $a_3$. This local adaption was also observed when travelers move from $a_4$ to $a_6$, instead of $a_7$. As a result, path 5 got the highest probability ($P_5 = 0.434$), followed by path 2 ($P_2 = 0.268$), path 4 ($P_4 = 0.160$), path 1 ($P_4 = 0.099$) and path 3 ($P_4 = 0.040$). This result implies that travelers cannot take the globally optimal path when some attributes are only locally perceived by them. 
As such, our model can describe both global preferences for and local responses to network attributes through the specifications of decomposed global and local utility functions.

%%%%%%%%%%%%%%%%%%%%%%%%%%%%%%%%%%%%%
%%%Table of Case settings
%%%%%%%%%%%%%%%%%%%%%%%%%%%%%%%%%%%%%
\begin{table}[t]
	\centering 
	\footnotesize
	\caption{Three tested cases. The columns for $x_1$ and $x_2$ indicate whether the attributes are global or local variables and their coefficients in the utility function. The resultant global and local utilities $v_G, v_L$ for each case are shown in the fourth and fifth columns.}
	\label{tb:cases}
	\begin{tabular*}{\hsize}{@{\extracolsep{\fill}}ccccc@{}}
		\toprule
		Case & $x_1$ & $x_2$ & $v_G$ & $v_L$ \\
		\midrule
		1 & Global, $-1$ & - & $-x_1$ & $0$ \\
		2 & Global, $-1$ & Global, $-1$ & $-(x_1+x_2)$ & $0$ \\
		3 & Global, $-1$ & Local, $-1$ & $-x_1$ & $-x_2$ \\
		\bottomrule
	\end{tabular*}
\end{table}
%%%%%%%%%%%%%%%%%%%%%%%%%%%%%%%%%%%%%

%%%%%%%%%%%%%%%%%%%%%%%%%%%%%%%%%%%%%
%%%Table of example results
%%%%%%%%%%%%%%%%%%%%%%%%%%%%%%%%%%%%%
\begin{table}[t]
	\centering 
	\footnotesize
	\caption{Path probabilities for different cases.}
	\label{tb:probs}
	\begin{tabular*}{\hsize}{@{\extracolsep{\fill}}cccccc@{}}
		\toprule
		Case & $P_1$ & $P_2$ & $P_3$ & $P_4$ & $P_5$ \\
		\midrule
        1 & $0.183$ & $0.183$ & $0.498$ & $0.067$ & $0.067$ \\
        2 & $0.192$ & $0.521$ & $0.026$ & $0.070$ & $0.192$ \\
        3 & $0.099$ & $0.268$ & $0.040$ & $0.160$ & $0.434$ \\
		\bottomrule
	\end{tabular*}
\end{table}
%%%%%%%%%%%%%%%%%%%%%%%%%%%%%%%%%%%%%

%%%%%%%%%%%%%%%%%%%%%%%%%%%%%%%%%%%%%
%%%Figure of prob with different mu
%%%%%%%%%%%%%%%%%%%%%%%%%%%%%%%%%%%%%
\begin{figure}[t] %tb
	\begin{center}
		\includegraphics[scale=.6]{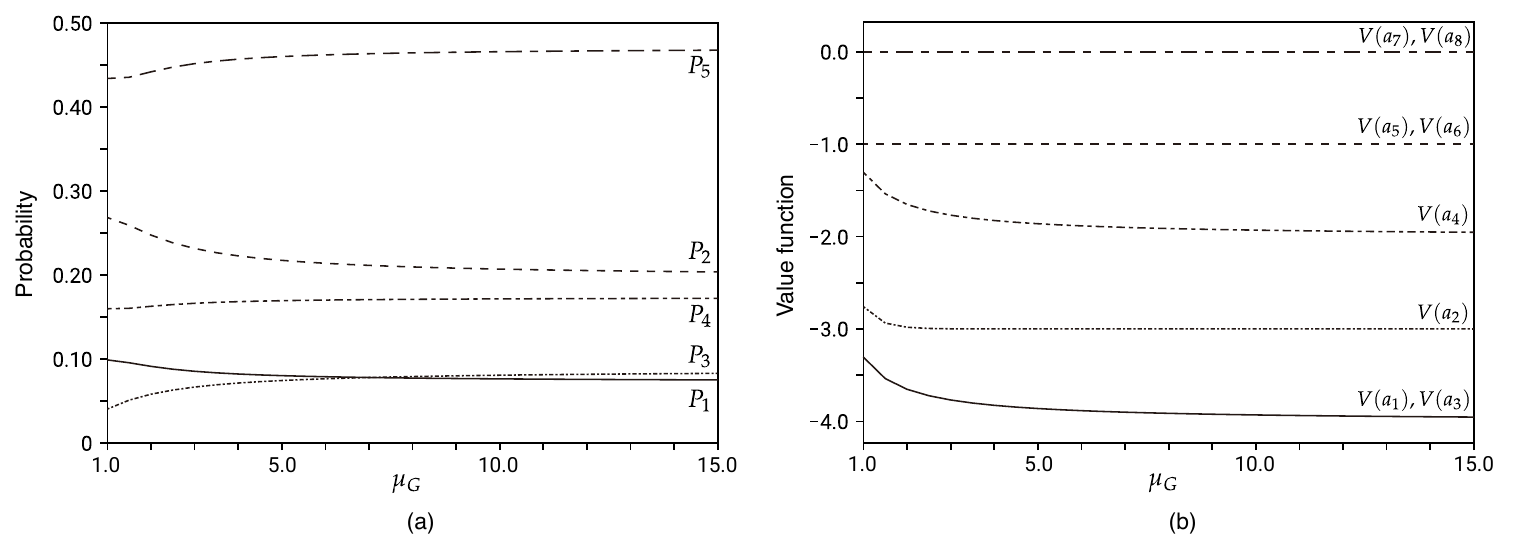}
		\caption{Chage in path probabilities (a) and value functions (b) with different $\mu_G$ values in case 3.}
		\label{fig:prob_mu} 
	\end{center}
\end{figure}
%%%%%%%%%%%%%%%%%%%%%%%%%%%%%%%%%%%%%

Next, to discuss the effect of the size of the scale parameter $\mu_G$, we computed the path probabilities with different values of $\mu_G$ where $\mu$ and $\gamma$ are both fixed to one, thereby $\gamma \mu / \mu_G = 1/\mu_G$. The results in Figure \ref{fig:prob_mu}(a) show that the probabilities gradually converge to certain values as $\mu_G$ grows. This is because, when $\mu_G$ goes to a sufficiently large value, the global path choice becomes deterministic, and the value function can be approximated by $V^d(k) \approx \max_{a \in A(k)} \{v_G(a|k) + V^d(a)\}$ describing the deterministic maximum path utility from link $k$ to $d$. This change in the value function is displayed in Figure \ref{fig:prob_mu}(b). Also, the difference between the values $V^d(a_1)$ and $V^d(a_2)$ of links $a_1$ and $a_2$ gets larger according to the increase in $\mu_G$. This explains that the probabilities of paths 1 and 2 whose first link is $a_1$ decrease, whereas those of paths 3-5 whose first link is $a_2$ increase. Note that although the certainty of travelers' perception of the global utility increases, they still locally maximize their utilities whose uncertainty is characterized by the scale $\mu$.

\section{Learning preferences from observed network paths}\label{sec:estimation}
In this section, we present the estimation of the proposed global-local path choice model based on maximum likelihood.
Unlike most of the previous adaptive (plan-action) path choice models, the proposed model can be estimated only with observations of paths that travelers actually took, without the information of \textit{plans}. 
Consider we have path observations $r_n = [a_1, \ldots, a_{J_n}]$, $n \in \{1, \ldots, N\}$, where an observed path $r_n$ is a sequence of links of length $J_n$, and the last element $a_{J_n}$ corresponds to its destination $d_n$. The log-likelihood function of the proposed model is
\begin{equation}
    \label{eq:LL}
    LL(\mathbold{\theta}; \mathbold{r}) = \sum^N_{n=1}\sum^{J_n-1}_{j=1} \ln p^{d_n}(a_{j+1}|a_j)
    = \sum^N_{n=1}\sum^{J_n-1}_{j=1} \left\{
    v(a|k) + V^d(a)
    - \ln \sum_{a' \in A(k)} e^{v(a'|k) + V^d(a')}
    \right\}
\end{equation}
where we assume a linear-in-parameters formulation of the reward functions and  consider $\mu$ and $\gamma$ to be one, and $\mathbold{\theta} = (\mathbold{\beta}_L, \mathbold{\beta}_G, \mu_G)$ are the parameters to be estimated. 

The maximum likelihood estimation is performed by \cite{Rust1987}'s nested fixed point (NFXP) algorithm, in which we iteratively solve the global value functions through the system of linear equation (\ref{eq:linearsystem}) and perform the outer loop nonlinear optimization based on the BFGS method. 
As the inner global value function computation was discussed in Section \ref{sec:valuefunc}, the rest of this section focuses on the derivation of the gradient of the likelihood function for the outer nonlinear optimization algorithm.
% in which we combine an outer nonlinear optimization BFGS to maximize the likelihood (\ref{eq:LL}) over parameter space and an inner algorithm to solve the global value function for each parameter. 
% As the global value function was discussed in Section \ref{sec:valuefunc}, the rest of this section focuses on the gradient of the likelihood function for the outer nonlinear optimization algorithm.
%This approach has also been applied in most recursive path choice models \citep{Fosgerau2013RL, Mai2015NRL, oyama2023capturing}.

% With $\mu$ fixed to one, the gradient with respect to a given parameter $\theta_i$ is
% \begin{align}
%     \label{eq:derivative}
%     \pdv{LL}{\theta_i} &= \pdv*{\sum^N_{n=1}\sum^{J_n-1}_{j=1} \left[ (v(a_{j+1}|a_j) + V^{d_n}(a_{j+1})) - \ln \sum_{a \in A(a_j)}  \exp (v(a|a_j) + V^{d_n}(a)) \right]}{\theta_i} \nonumber\\
%     &= \sum^N_{n=1}\sum^{J_n-1}_{j=1} \left[ \left(\pdv{v(a_{j+1}|a_j)}{\theta_i} + \pdv{V^{d_n}(a_{j+1})}{\theta_i}\right) - \sum_{a \in A(a_j)} p^{d_n}(a|a_j) \left(\pdv{v(a_{j+1}|a_j)}{\theta_i} + \pdv{V^{d_n}(a)}{\theta_i}\right)  \right].
% \end{align}

The gradient of (\ref{eq:LL}) with respect to each specific parameter is derived as (we omit the superscript for destination $d_n$ here for simplicity):
\begin{align}
    \label{eq:derivative_spec}
    \pdv{LL}{\beta^{L}_i} &= \sum^N_{n=1}\sum^{J_n-1}_{j=1} \left\{ x^{L,i}_{a_{j+1}|a_j} - \mathbb{E}_{\mathbf{p}} [\mathbf{x}^{L}_i ~|~ a_j]  \right\} \\ %\mathbb{E}_{\mathbf{p}} [\mathbf{x}^{L,i}_{\mathbf{a}|a_j}] \\
    \pdv{LL}{\beta^G_{h}} &= \sum^N_{n=1}\sum^{J_n-1}_{j=1} \left\{ x^{G,h}_{a_{j+1}|a_j} + \pdv{V(a_{j+1})}{\beta^G_h} - \mathbb{E}_{ \mathbf{p}} \left[\mathbf{x}^{G}_{h} + \pdv{\mathbf{V}}{\beta^G_h} ~\middle|~ a_j \right]  \right\} \\ %\mathbb{E}_{ \mathbf{p}} \left[\mathbf{x}^{G,h}_{\mathbf{a}|a_j} + \pdv{\mathbf{V(a)}}{\beta^G_h}\right]
    \pdv{LL}{\mu_{G}} &= \sum^N_{n=1}\sum^{J_n-1}_{j=1} \left\{ \pdv{V(a_{j+1})}{\mu_G} - \mathbb{E}_{\mathbf{p}} \left[\pdv{\mathbf{V}}{\mu_G} ~\middle|~ a_j\right]  \right\} %\mathbb{E}_{\mathbf{p}} \left[\pdv{\mathbf{V(a)}}{\mu_G}\right]
\end{align}
where $\mathbb{E}_{\mathbf{p}}[\mathbf{x} ~|~ k] \equiv \sum_{a\in A(k)}p(a|k)x_{a|k}$, the expected value of $\mathbf x$ according to the action probability $\mathbf p$ conditional on the state $k$. 

To compute the above gradients, we still need the gradients of the global value function $V$ with respect to $\beta_G$ and $\mu_G$, which are
\begin{align}
    \label{eq:pdv_V_b}
    \pdv{\mathbf{V}}{\beta^G_h} &= (\mathbf{I} - \mathbf{P}^\top_G)^{-1} \mathbf{D}^G_h \\ %\mathbb{E}_{ \mathbf{P}_G}\left[\mathbf{x}^{G,h}\right]
    \label{eq:pdv_V_mu}
    \pdv{\mathbf{V}}{\mu_G} &= - \frac{1}{\mu_G^2} (\mathbf{I} - \mathbf{P}^\top_G)^{-1} \mathbf{H}_G
\end{align}
%$P_G(a|k) = M_{ka}z_a/z_k$
where $P_G(a|k) = M_{ka}z_a/z_k$ is the global link choice probability matrix that is consistent with the RL models \citep{Fosgerau2013RL}; $D^G_{h}(k) = \sum_{a\in A(k)} P_G(a|k) x^{G,h}_{a|k}$ is the expected value of the $h$-th global instantaneous variable in state $k$; and $H_G(k) = - \sum_{a \in A(K)} P_G(a|k) \ln P_G(a|k)$ is the global path choice entropy function. Because $(\mathbf{I} - \mathbf{P}^\top_G)$ is invertible \citep{Baillon2008MCA, Fosgerau2013RL}, the gradients of the value function (\ref{eq:pdv_V_b}) and (\ref{eq:pdv_V_mu}) are computed by solving the systems of linear equations.

More details of the derivation are provided in \ref{app:gradient}.

\section{Numerical results}\label{sec:result}
This section presents several numerical results of the estimation of the global-local path choice model. We first show an experiment using synthetic data to examine the parameter reproducibility between models with different assumptions on the traveler's perception of a network attribute. We then provide a real application result in the case study of pedestrian path choice, where we introduce an attribute of visual street quality extracted from street images. 

The model estimation was performed by the NFXP algorithm as discussed in Section \ref{sec:estimation}.
% For the model estimation, we used the nested fixed point (NFXP) algorithm \citep{Rust1987}, in which we solve the global value functions through the system of linear equation (\ref{eq:linearsystem}) and perform the outer loop nonlinear optimization based on the BFGS method. 
In addition, we used \cite{oyama2023capturing}'s two-phase estimation procedure: we first estimated a prism-constrained RL model \citep{oyama2019prism} only with the global utility $u_G$, whose estimates were then used as the starting point for the estimation of the proposed model where the initial values of the coefficients of local attributes were set to zero. %$\beta^{(0)}_G$, $\beta^{(0)}_L$ 
This procedure allowed us to mitigate the numerical issue regarding the evaluation of the global value functions during the estimation \citep[please refer to][for the detail]{oyama2023capturing}.
The standard error of the estimates and the confidence intervals of indicators were calculated using bootstrapping with 100 iterations. %in the real application, 
We implemented modeling, estimation, and simulation by writing our own Python code\footnote{The code will be made publicly available after publication.}.
%, on a machine with 14 cores Intel Xeon W processors (2.5 GHz) and 64 GB of RAM. %which will be available online and upon request

\subsection{Experiment with simulated observations}\label{sec:synthetic}
We first show the result of an experiment using synthetic data in the Sioux-Falls network \citep{SiouxFalls2016}. In this experiment, we focus on a specific attribute and compare models that differently assume the perception of the attribute.
More specifically, we compare the following specifications of the instantaneous reward function:
\begin{subequations} \label{eq:pputil}
\begin{align}\centering
	&\left\{
    \begin{array}{c l}
    v_G(a|k) &=   (\beta^G_{\rm{len}} + \beta^G_{\rm{cap}}x^{\rm cap}_a)x^{\rm len}_a    - 20 x^{\rm uturn}_{a|k}\\
    v_L(a|k) &= 0\\
    \end{array}
    \right. \label{eq:SFglobal}\\
	&\left\{
    \begin{array}{c l}
    v_G(a|k) &=   \beta^G_{\rm{len}} x^{\rm len}_a - 20 x^{\rm uturn}_{a|k}\\
    v_L(a|k) &=  \beta^L_{\rm{cap}} x^{\rm cap}_a x^{\rm len}_a\\
    \end{array}
    \right. \label{eq:SFlocal}\\
    &\left\{
    \begin{array}{c l}
    v_G(a|k) &=   (\beta^G_{\rm{len}} + \beta^G_{\rm{cap}}x^{\rm cap}_a)x^{\rm len}_a  - 20 x^{\rm uturn}_{a|k}\\
    v_L(a|k) &=  \beta^L_{\rm{cap}} x^{\rm cap}_a x^{\rm len}_a\\
    \end{array}
    \right. \label{eq:SFglobalocal}
\end{align}
\end{subequations}
where $x^{\rm len}_a$ is the length of link $a$, and $x^{\rm cap}_a$ is its capacity divided by the maximum link capacity in the network, whose effect is captured by an interaction with the link length to be consistent with the link-additive assumption. % \citep[see][for more details]{oyama2023capturing}
The different specifications respectively assume that the attribute $x^{\rm cap}_a$ is globally perceived (\ref{eq:SFglobal}), only locally perceived (\ref{eq:SFlocal}), and both globally and locally perceived (\ref{eq:SFglobalocal}), which we name Model G, Model L, and Model GL respectively. Note that Model G corresponds to the RL (global path choice) model \citep{Fosgerau2013RL}.

For the experiment, we set the true parameters to $(\beta^G_{\rm{len}}, \beta^G_{\rm{cap}}, \beta^L_{\rm{cap}}) = (-2.5, 0.5, 2.0)$ and generate two different synthetic datasets from Model G and Model L by implementing Monte Carlo simulations. For each model, we generated 1,000 path observations for each of 24 OD pairs (i.e., 24,000 in total), and split them into 10 samples, each of which thus had 2,400 observations. %Note that we did not observe any path with a loop where a repeated link is considered a loop.
Using these two different synthetic datasets generated from Model G and Model L (we call Data G and Data L), we estimated all three models, where $\mu$, $\mu_G$ and $\gamma$ are fixed to one. 
The results are reported in Table \ref{tb:SFestRes}. 

As expected, when the estimated model had the same specification as the model used to generate data, the estimation well reproduced the true parameters: on average over 10 samples, the estimates of Model G with Data G were $(\hat{\beta}^G_{\rm{len}}, \hat{\beta}^G_{\rm{cap}}) = (-2.49, 0.51)$, and those of Model L with Data L were $(\hat{\beta}^G_{\rm{len}}, \hat{\beta}^L_{\rm{cap}}) = (-2.47, 1.97)$. 
In contrast, when estimating a different model to the model used for data generation (the estimation of Model G with Data L and that of Model L with Data G), the estimated parameters included biases and were significantly different from the true values. 

As for the estimation of Model GL that introduced the attribute $x^{\rm cap}_a$ to both global and local utilities, the coefficients of effects that were not included in the model for data generation were estimated as not significantly different from zero ($\hat{\beta}^L_{\rm{cap}} = -0.05$ for Data G, and $\hat{\beta}^G_{\rm{cap}} = 0.03$ for Data L), while the other parameters were reproduced well. 

These results show the difference between the global and local effects of an attribute, which our model can capture by flexibly defining the global and local utility functions. Moreover, it was also shown that we can analyze to what extent and which attributes globally and locally affect the path choice behavior by estimating and comparing different specifications with respect to attributes of interest, including one with the attribute introduced to both the global and local utilities. 
%well reproduced

%%%%%%%%%%%%%%%%%%%%%%%%%%%%%%%%%%%%%
%%%Table of estimation results
%%%%%%%%%%%%%%%%%%%%%%%%%%%%%%%%%%%%%
\begin{table}[t]
	\centering 
	\footnotesize
	\caption{Estimation results: averages and standard errors of the estimates over 10 samples.}
	\label{tb:SFestRes}
	\begin{tabular*}{\hsize}{@{\extracolsep{\fill}}llrrrrrrr@{}}
		\toprule
        && \multicolumn{7}{c}{Estimated model}\\
        \cmidrule(lr){3-9}
        && \multicolumn{2}{c}{Model G (\ref{eq:SFglobal})}  & \multicolumn{2}{c}{Model L (\ref{eq:SFlocal})}  & \multicolumn{3}{c}{Model GL (\ref{eq:SFglobalocal})}  \\
        \cmidrule(r){3-4}\cmidrule(lr){5-6}\cmidrule(l){7-9}
        Data generated by &  & $\hat{\beta}^G_{\rm{len}}$ & $\hat{\beta}^G_{\rm{cap}}$ & $\hat{\beta}^G_{\rm{len}}$ & $\hat{\beta}^L_{\rm{cap}}$ & $\hat{\beta}^G_{\rm{len}}$ & $\hat{\beta}^G_{\rm{cap}}$ & $\hat{\beta}^L_{\rm{cap}}$ \\
        \midrule
        Model G (\ref{eq:SFglobal}) & Average & -2.49 & 0.51 & -1.87 & 0.10 & -2.50 & 0.52 & -0.05 \\
        (Data G) & Std.err. & 0.22 & 0.10 & 0.09 & 0.09 & 0.22 & 0.10 & 0.06 \\
        % & TRUE & -2.50 & 0.50 & -2.50 & 2.00 & -2.50 & 0.50 & 0.00 \\
        \midrule
        Model L (\ref{eq:SFlocal}) & Average & -2.72 & 0.78 & -2.47 & 1.97 & -2.48 & 0.03 & 1.94 \\
        (Data L) & Std.err. & 0.14 & 0.08 & 0.10 & 0.13 & 0.13 & 0.11 & 0.12 \\
        \bottomrule
	\end{tabular*}
\end{table}
%%%%%%%%%%%%%%%%%%%%%%%%%%%%%%%%%%%%%

\subsection{Real pedestrian path choice application}\label{sec:application}
\begin{figure}[t]
	\begin{center}
        \includegraphics[width=15cm]{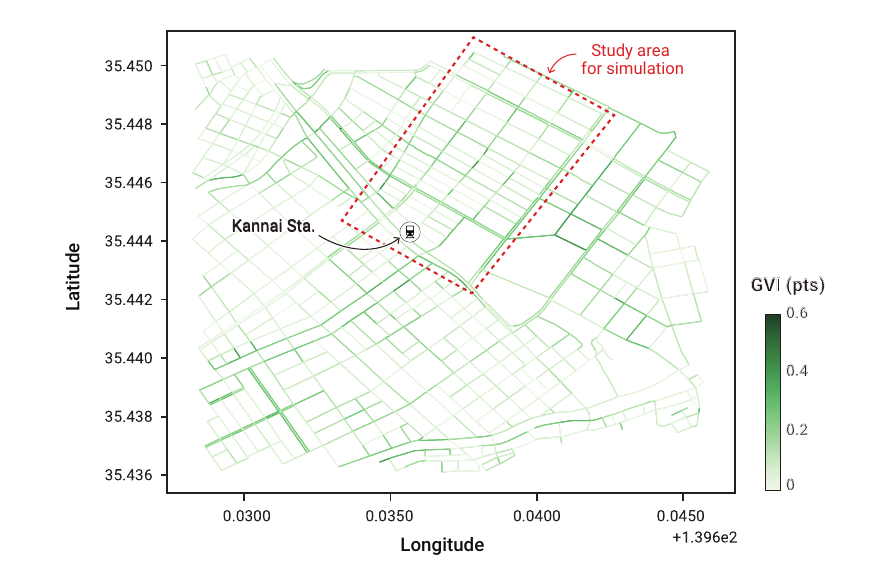}
		\caption{Pedestrian network for real application: A mile square centered on the Kannai station. Deeper colors indicate higher GVI values of the streets. The area for simulation study is enclosed by the red-dotted rectangle.}
		\label{fig:ppnet} 
	\end{center}
\end{figure}

We then show an application of the proposed path choice model to pedestrian path choice analysis using real path observations. Because walking is a slow mode of transportation, pedestrians may visually perceive the street environment while walking and locally adjust their path choice behavior, which we analyze by using the proposed model. 
The data is the same as used in \cite{oyama2023capturing}, based on GPS trajectories collected through a complementary survey of the Sixth Tokyo Metropolitan Region Person Trip Survey \citep{TokyoPT}, in the Kannai area, Yokohama city, Japan. The pedestrian network for the case study contains 724 nodes and 2398 links with 8434 link pairs, as shown in Figure \ref{fig:ppnet}.

%%%%%%%%%%%%%%%%%%%%%%%%%%%%%%%%%%%%%
%%%Figure of Validation results
%%%%%%%%%%%%%%%%%%%%%%%%%%%%%%%%%%%%%
%%%%%%%%%%%%%%%%%%%%%%%%%%%%%%%%%%%%%

In this case study, we consider the following three specifications of the reward function:
\begin{subequations} \label{eq:pputil}
\begin{align}\centering
	&\left\{
    \begin{array}{c l}
    v_G(a|k) &=   (\beta^G_{\rm{len}} + \beta^G_{\rm{walk}} x^{\rm walk}_a   + \beta^G_{\rm{green}} x^{\rm green}_a) x^{\rm len}_a  
    + \beta^G_{\rm{cross}} x^{\rm cross}_a - 20 x^{\rm uturn}_{a|k}\\
    v_L(a|k) &= 0\\
    \end{array}
    \right. \label{eq:global_model}\\
	&\left\{
    \begin{array}{c l}
    v_G(a|k) &=   (\beta^G_{\rm{len}} + \beta^G_{\rm{walk}} x^{\rm walk}_a) x^{\rm len}_a  
    + \beta^G_{\rm{cross}} x^{\rm cross}_a - 20 x^{\rm uturn}_{a|k}\\
    v_L(a|k) &=  \beta^L_{\rm{green}} x^{\rm green}_a x^{\rm len}_a\\
    \end{array}
    \right. \label{eq:local_model}\\
    &\left\{
    \begin{array}{c l}
    v_G(a|k) &=   (\beta^G_{\rm{len}} + \beta^G_{\rm{walk}} x^{\rm walk}_a   + \beta^G_{\rm{green}} x^{\rm green}_a) x^{\rm len}_a  
    + \beta^G_{\rm{cross}} x^{\rm cross}_a - 20 x^{\rm uturn}_{a|k}\\
    v_L(a|k) &=  \beta^L_{\rm{green}} x^{\rm green}_a x^{\rm len}_a\\
    \end{array}
    \right. \label{eq:globalocal_model}
\end{align}
\end{subequations}
where $x^{\rm len}_a$ and $x^{\rm walk}_a$ are respectively the length and sidewalk width (m/10) of link $a$; $x^{\rm cross}_a$ is the dummy variable of $a$ being a crosswalk; $x^{\rm green}_a$ is the green view index (GVI) of the street, extracted as the vegetation pixel ratio ($\in [0, 1]$) from Google Street View images using a deep learning model\footnote{\cite{oyama2023capturing} assumed $x^{\rm green}_a$ to be a dummy variable simply representing the presence of streetscape greenery. Instead, we calculated the GVI through semantic segmentation by DeepLabv3 \citep{deeplabv3plus2018}.}. 
We capture the effects of sidewalk widths and GVI by interactions with the length so that the link-additive nature of the global path choice is retained.
We also add a fixed negative u-turn penalty ${\rm Uturn}_{a|k}$, following previous studies of RL models \citep[e.g.,][]{Fosgerau2013RL}. The parameters to be estimated are $\beta^G_{\rm{len}}$, $\beta^G_{\rm{walk}}$, $\beta^G_{\rm{cross}}$, $\beta^G_{\rm{green}}$ and $\beta^L_{\rm{green}}$. 

We hypothesized that the visual quality of the streets like GVI is locally perceived by pedestrians and affects their local path choice. To test this hypothesis, we compare the three specifications (\ref{eq:global_model})-(\ref{eq:globalocal_model}). The first specification (\ref{eq:global_model}) assumes the GVI as a global attribute, which coincides with an RL model \citep{Fosgerau2013RL}, and the second (\ref{eq:local_model}) does it as a local attribute. 
The third specification  (\ref{eq:globalocal_model}) introduces the attribute to both global and local utilities, which is considered following the suggestion from the experiment in Section \ref{sec:synthetic}. 
In addition, we compare the results with and without $\mu_G$ estimated for the specifications (\ref{eq:local_model}) and (\ref{eq:globalocal_model}).

In summary, we estimate the following five models:
\begin{itemize}
    \item \textbf{Model 1}: A global path choice model (\ref{eq:global_model}) which coincides with an RL model
    \item \textbf{Model 2}: A global-local path choice model (\ref{eq:local_model}) with $\mu_G$ fixed to one
    \item \textbf{Model 3}: A global-local path choice model (\ref{eq:local_model}) with $\mu_G$ to be estimated
    \item \textbf{Model 4}: A global-local path choice model (\ref{eq:globalocal_model}) with $\mu_G$ fixed to one
    \item \textbf{Model 5}: A global-local path choice model (\ref{eq:globalocal_model}) with $\mu_G$ to be estimated
\end{itemize}
where $\mu$ and $\gamma$ are fixed to one for all the models\footnote{The results for the cases with $\gamma < 1$ are reported in \ref{app:discount}.}. 
Note that in theory any attributes can be introduced into both global and local utility functions, but we focus our interest on the GVI attribute in this case study for brevity and interpretation.

%For the model estimation, we used the nested fixed point (NFXP) algorithm \citep{Rust1987} where the global value functions are solved through the system of linear equation (\ref{eq:linearsystem}) and the outer loop nonlinear optimization is performed using the BFGS method. 
% The standard error of the estimates and the confidence intervals of indicators are calculated using bootstrapping with 100 iterations. Moreover, we used \cite{oyama2023capturing}'s two-phase estimation procedure; first estimate a prism-constrained RL model \citep{oyama2019prism} only with the global utility $u_G$, whose estimates are then used as the starting point  $\beta^{(0)}_G$ for the estimation of the proposed model where the initial values of the coefficients of local attributes $\beta^{(0)}_L$ are set to zero. 
% This procedure allows us to mitigate the numerical issue regarding the evaluation of the global value functions \citep[see][for the detail]{oyama2023capturing}.

\subsubsection{Estimation results} 
The model estimation results are reported in Tables \ref{tb:estimation_res1} and \ref{tb:estimation_res2}. %, where the discount factor $\gamma$ was fixed to one 
For all the models, we obtained the expected signs of the parameters, and most of them were statistically and significantly different from the references. %(one for $\mu_G$ and zero for the others)
From the signs of the estimates, we generally found that pedestrians have global preferences to walk along paths with shorter lengths, less number of crosswalks, and wider sidewalks. As for the GVI, its positive signs indicate the positive effect of streetscape greenery on pedestrian path choice. 

Models 2 and 3 which introduced the GVI as a local attribute obtained a higher goodness-of-fit than Model 1 which introduced it as a global attribute.
The estimate $\hat{\beta}^L_{\rm green}$ for Models 2 and 3 was statistically and significantly different from zero, while that for Model 1 $\hat{\beta}^G_{\rm green}$ was not. These results suggest that the volume of streetscape greenery affects the local responses of pedestrians rather than their global path choices.
%pedestrians locally react to the visual greenery on the streets, rather than they globally perceive it.
%Note that the estimate $\beta^G_{\rm green}$ for the global model was not different from zero with statistical significance, while that for the local model $\beta^L_{\rm green}$ was statistically and significantly different from zero.

The scale of the global value function in Model 3 was estimated as $\hat{\mu}_G = 1.280$, indicating higher certainty of the pedestrians' perception of the global utilities compared to the local utilities. The likelihood ratio test between Models 2 and 3 ($\chi^2 = 15.46$) also indicates that Model 3 better fits the data than Model 2.

The estimation results of Models 4 and 5, which introduced the GVI attribute into both global and local utilities, further support the fact that pedestrians locally react to the volume of streetscape greenery. In these models, the estimate $\hat{\beta}^G_{\rm green}$ of GVI as a global attribute was not statistically and significantly different from zero. 
The likelihood ratio tests between Models 2 and 4 ($\chi^2 = 0.15$) and between Models 3 and 5 ($\chi^2 = 0.12$), together with their AIC values, also suggest that the additional introduction of GVI to the global utility did not contribute to a significant improvement of the model fit. 
Therefore, we focus on the comparison between Models 1-3 in the following discussion. 
%conclude that the volume of streetscape greenery affects the local responses of pedestrians rather than their global path choices, and 

%%%%%%%%%%%%%%%%%%%%%%%%%%%%%%%%%%%%%
%%%Table of estimation results
%%%%%%%%%%%%%%%%%%%%%%%%%%%%%%%%%%%%%
\begin{table}[t]
	\centering 
	\footnotesize
	\caption{Estimation results of Models 1-3.}
	\label{tb:estimation_res1}
	\begin{tabular*}{\hsize}{@{\extracolsep{\fill}}ccrrrrrrrrr@{}}
		\toprule
        && \multicolumn{3}{c}{Model 1} & \multicolumn{3}{c}{Model 2} & \multicolumn{3}{c}{Model 3} \\ \cmidrule(lr){3-5}\cmidrule(lr){6-8}\cmidrule(lr){9-11}
	    & Parameter & Estimate & std. err. & t-stat$^\dag$ & Estimate & std. err. & t-stat & Estimate & std. err. & t-stat \\
        \midrule
        Global & $\hat{\beta}_{\rm len}$ & -0.322 & 0.011 & -29.65$^{***}$ & -0.316 & 0.010 & -31.30$^{***}$ & -0.290 & 0.016 & -18.62$^{***}$ \\
         & $\hat{\beta}_{\rm cross}$ & -0.927 & 0.055 & -16.71$^{***}$ & -0.886 & 0.057 & -15.51$^{***}$ & -0.816 & 0.066 & -12.35$^{***}$ \\
         & $\hat{\beta}_{\rm walk}$ & 0.063 & 0.010 & 6.28$^{***}$ & 0.069 & 0.009 & 7.64$^{***}$ & 0.062 & 0.010 & 6.29$^{***}$ \\
         & $\hat{\beta}_{\rm green}$ & 0.072 & 0.057 & 1.26$^{~~~}$ & - & - & - & - & - & - \\
        Local & $\hat{\beta}_{\rm green}$ & - & - & - & 0.139 & 0.046 & 3.04$^{***}$ & 0.096 & 0.044 & 2.18$^{**~}$ \\
        Scale & $\hat{\mu}_G$ & Fixed & - & - & Fixed & - & - & 1.280 & 0.142 & 1.98$^{**~}$ \\
        \midrule
        \multicolumn{2}{l}{Path observations} &  &  & 410 &  &  & 410 &  &  & 410 \\
        \multicolumn{2}{l}{Log-likelihood} &  &  & -1701.2 &  &  & -1697.3 &  &  & -1689.6 \\
        \multicolumn{2}{l}{AIC} &  &  & 3410.4 &  &  & 3402.6 &  &  & 3389.2 \\
		\bottomrule
        \multicolumn{11}{l}{$^\dag$ Confidence level of statistical significance: $^{***}$: $p \le 0.01$; $^{**}$: $p \in (0.01, 0.05]$; $^{*}$: $p \in (0.05, 0.1]$}
	\end{tabular*}

     \caption{Estimation results of Models 4-5.}
    \label{tb:estimation_res2}
    \begin{tabular*}{\hsize}{@{\extracolsep{\fill}}ccrrrrrr@{}}
        \toprule
        && \multicolumn{3}{c}{Model 4} & \multicolumn{3}{c}{Model 5} \\ \cmidrule(lr){3-5}\cmidrule(lr){6-8}
        & Parameter & Estimate & std. err. & t-stat$^\dag$ & Estimate & std. err. & t-stat  \\
        \midrule
        Global & $\hat{\beta}_{\rm len}$ &
        -0.317 & 0.011 & -27.85$^{***}$ & -0.291 & 0.017 & -17.25$^{***}$ \\
         & $\hat{\beta}_{\rm cross}$ & 
         -0.888 & 0.058 & -15.34$^{***}$ & -0.817 & 0.067 & -12.20$^{***}$ \\
         & $\hat{\beta}_{\rm walk}$ & 
         0.067 & 0.010 & 6.49$^{***}$ & 0.060 & 0.010 & 5.85$^{***}$ \\
         & $\hat{\beta}_{\rm green}$ & 
         0.019 & 0.066 & 0.28$^{~~~}$ & 0.016 & 0.062 & 0.26$^{~~~}$ \\
        Local & $\hat{\beta}_{\rm green}$ & 
        0.132 & 0.054 & 2.42$^{**~}$ & 0.090 & 0.050 & 1.79$^{*~~}$ \\
        Scale & $\hat{\mu}_G$ & Fixed
         & - & - & 1.280 & 0.142 & 1.97$^{**~}$ \\
        \midrule
        \multicolumn{2}{l}{Path observations} &  &  & 410 &  &  & 410  \\
        \multicolumn{2}{l}{Log-likelihood} &  &  & -1697.3 &  &  & -1689.5 \\
        \multicolumn{2}{l}{AIC} &  &  & 3404.6 &  &  & 3391.1 \\
        \bottomrule
        \multicolumn{8}{l}{$^\dag$ Confidence level of statistical significance: $^{***}$: $p \le 0.01$; $^{**}$: $p \in (0.01, 0.05]$; $^{*}$: $p \in (0.05, 0.1]$}
    \end{tabular*}
\end{table}
%%%%%%%%%%%%%%%%%%%%%%%%%%%%%%%%%%%%%

\subsubsection{Cross validation} 
We performed 20-fold cross-validation to compare the models with respect to out-of-sample prediction performance. For each dataset, the observations were split into estimation and holdout (validation) samples with a ratio of 80\% and 20\%. %, and prepared 20 sets of them 
The model performance was evaluated based on the log-likelihood obtained by applying the estimated model to the holdout sample. We computed the validation log-likelihood divided by the number of paths $LL_{i} = LL(\hat{\mathbold{\theta}}_{i}; \sigma_{i})/N_{i}$ for each holdout sample $i$ and then computed its average over samples $\overline{LL}_{i} = \frac{1}{p} \sum_{i=1}^{p} LL_{i}$, $\forall p \in \{1, \ldots, 20\}$. 

\begin{figure}[t]
	\begin{center}
        \includegraphics[width=15cm]{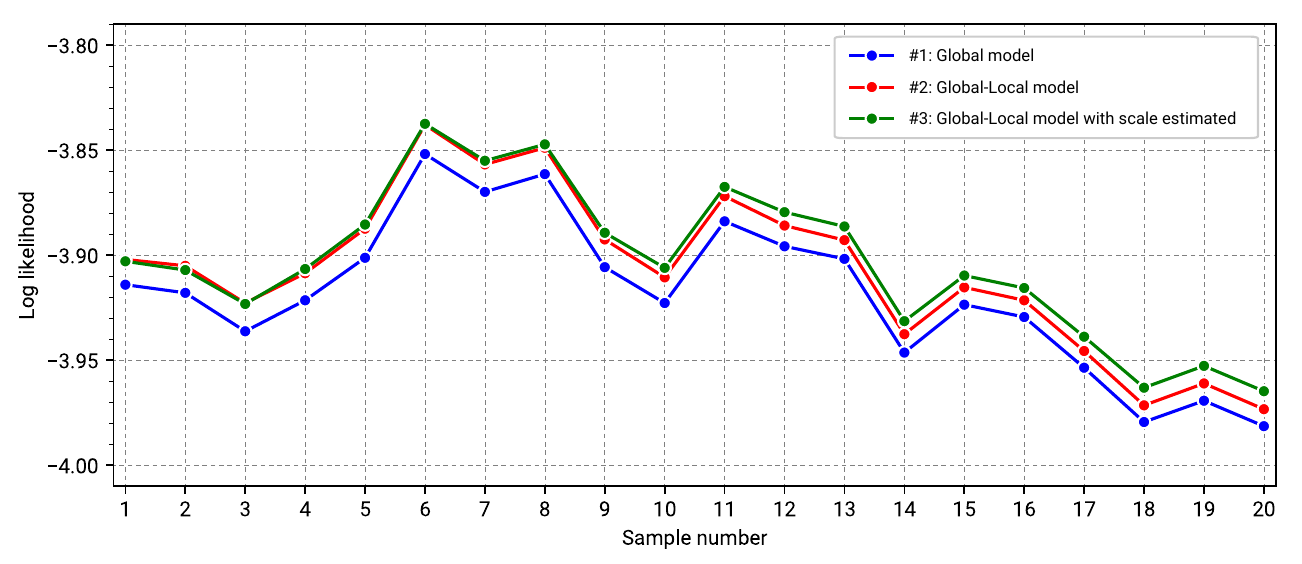}
		\caption{Validation results. The larger values (the upper positions) indicate better model prediction performance.}
		\label{fig:validation} 
	\end{center}
\end{figure}

Figure \ref{fig:validation} shows the validation results, and Table \ref{tb:validation} reports the average of the validation log-likelihood values over 20 holdout samples $\overline{LL}$ ($=\overline{LL}_{20}$). 
Models 2 and 3, which introduced the GVI attribute as a local attribute, got higher prediction performance than the global model (Model 1). This result suggests that capturing the effect of the volume of streetscape greenery on local responses better predicted the pedestrians' path choices than capturing its effect on global path preferences. 
%pedestrians locally perceive the visual quality of streets like GVI and adapt their path choice to the local context/environment. 

Although Model 3 had a better result than Model 2 on average, the improvement was relatively slight, and for some samples, the additional estimation of scale $\mu_G$ did not improve the out-of-sample prediction performance.

%%%%%%%%%%%%%%%%%%%%%%%%%%%%%%%%%%%%%
%%%Figure of Validation results
%%%%%%%%%%%%%%%%%%%%%%%%%%%%%%%%%%%%%
%%%%%%%%%%%%%%%%%%%%%%%%%%%%%%%%%%%%%

%%%%%%%%%%%%%%%%%%%%%%%%%%%%%%%%%%%%%
%%%Table of validation likelihood
%%%%%%%%%%%%%%%%%%%%%%%%%%%%%%%%%%%%%
\begin{table}[t]
	\centering 
	\footnotesize
	\caption{Average of validation log-likelihood values over 20 holdout samples.}
	\label{tb:validation}
	\begin{tabular*}{0.75\hsize}{@{\extracolsep{\fill}}cccc@{}}
		\toprule
		& Model 1 & Model 2 & Model 3~~~ \\
		\midrule
		~~~$\overline{LL}$ & -3.981 & -3.973 & -3.965~~~\\
		\bottomrule
	\end{tabular*}
\end{table}
%%%%%%%%%%%%%%%%%%%%%%%%%%%%%%%%%%%%%

\subsubsection{Willingness-to-walk}
We then analyze how changes in network attributes affect pedestrians' walking behavior using the estimated models. We first calculated \textit{willingness-to-walk} (WTW) measures of the network attributes by taking the ratio between an attribute of interest and the link length \citep{basu2022street}. 
Note that although our path choice model is a link-based model, it respects the link-additive nature of the global utility so that the sum of elemental link utilities of a path yields the global path utility, and the trade-off between the link-based variables represents the corresponding WTW measures of path-based models. 
%while \cite{basu2022street} which estimated path-based models. 

Table \ref{tb:wtw} reports the estimated WTW values. Each number represents the change in WTW, how many additional meters a pedestrian is willing to walk for every 100 meters of walking distance, caused by the unit change in the attribute of interest. 
Similar values were obtained among the different models. The WTW with respect to the crosswalk attribute indicates that each additional crosswalk along the path reduces the WTW by 28.0--28.7 meters per 100 meters on average. 
In contrast, a one-meter increase in sidewalk width along the path leads to an increase in pedestrians' WTW by 20.1--22.4 meters on average. The increase in GVI also has a positive effect on WTW: its change according to a 10\% increase in GVI was estimated on average 2.1 meters by the global model and 3.7--4.0 meters by Models 2 and 3.
%the increases in sidewalk width and GVI lead to an increase in pedestrians' WTW. 
% Although similar values were obtained among the different models, 

The confidence interval of the local WTW with respect to GVI increase was wider for Model 3 than that for Model 2, due to the scaling by $\hat{\mu}_G$. Because $\hat{\mu}_G$ was estimated at 1.280, Model 3 describes higher certainty in pedestrians' perception of the global utility than Model 2, and the WTW for the local attribute became less certain.

%%%%%%%%%%%%%%%%%%%%%%%%%%%%%%%%%%%%%
%%%Table of willingness-to-walk
%%%%%%%%%%%%%%%%%%%%%%%%%%%%%%%%%%%%%
\begin{table}[t]
	\centering 
	\footnotesize
	\caption{Estimated willingness-to-walk measures in meters per 100 meters walking distance.}
	\label{tb:wtw}
	\begin{tabular*}{\hsize}{@{\extracolsep{\fill}}lcccccc@{}}
		\toprule
        & \multicolumn{2}{c}{Model 1} & \multicolumn{2}{c}{Model 2} & \multicolumn{2}{c}{Model 3} \\ \cmidrule(lr){2-3}\cmidrule(lr){4-5}\cmidrule(lr){6-7}
        Variable & Mean & CI$^\dag$ & Mean & CI & Mean & CI \\
        \midrule
		One extra crosswalk along path & -28.7 & [-31.6, -25.5] & -28.0 & [-31.1, -24.9] & -28.2 & [-31.5, -25.0] \\
        One meter increase in sidewalk width & 20.1 & [15.5, 24.4] & 22.4 & [17.6, 26.3] & 21.7 & [17.0, 26.0] \\
        10\% (0.1 pts) increase in GVI (Global) & 2.07 & [0.92, 4.65] & - & - & - & - \\
        10\% (0.1 pts) increase in GVI (Local) & - & - &  3.95 & [1.38, 6.29] & 3.65 & [0.56, 6.65] \\
        % 10\% increase in GVI (Local) &  &  &  &  & 3.65 & (0.56, 6.65) \\
		\bottomrule
        \multicolumn{7}{l}{$^\dag$ 95\% confidence interval, calculated using bootstrapping with 100 iterations}
	\end{tabular*}
\end{table}
%%%%%%%%%%%%%%%%%%%%%%%%%%%%%%%%%%%%%

\subsubsection{Simulation with streetscape greenery increase policy}
\begin{figure}[H] %tb
	\begin{center}
		\includegraphics[scale=.5]{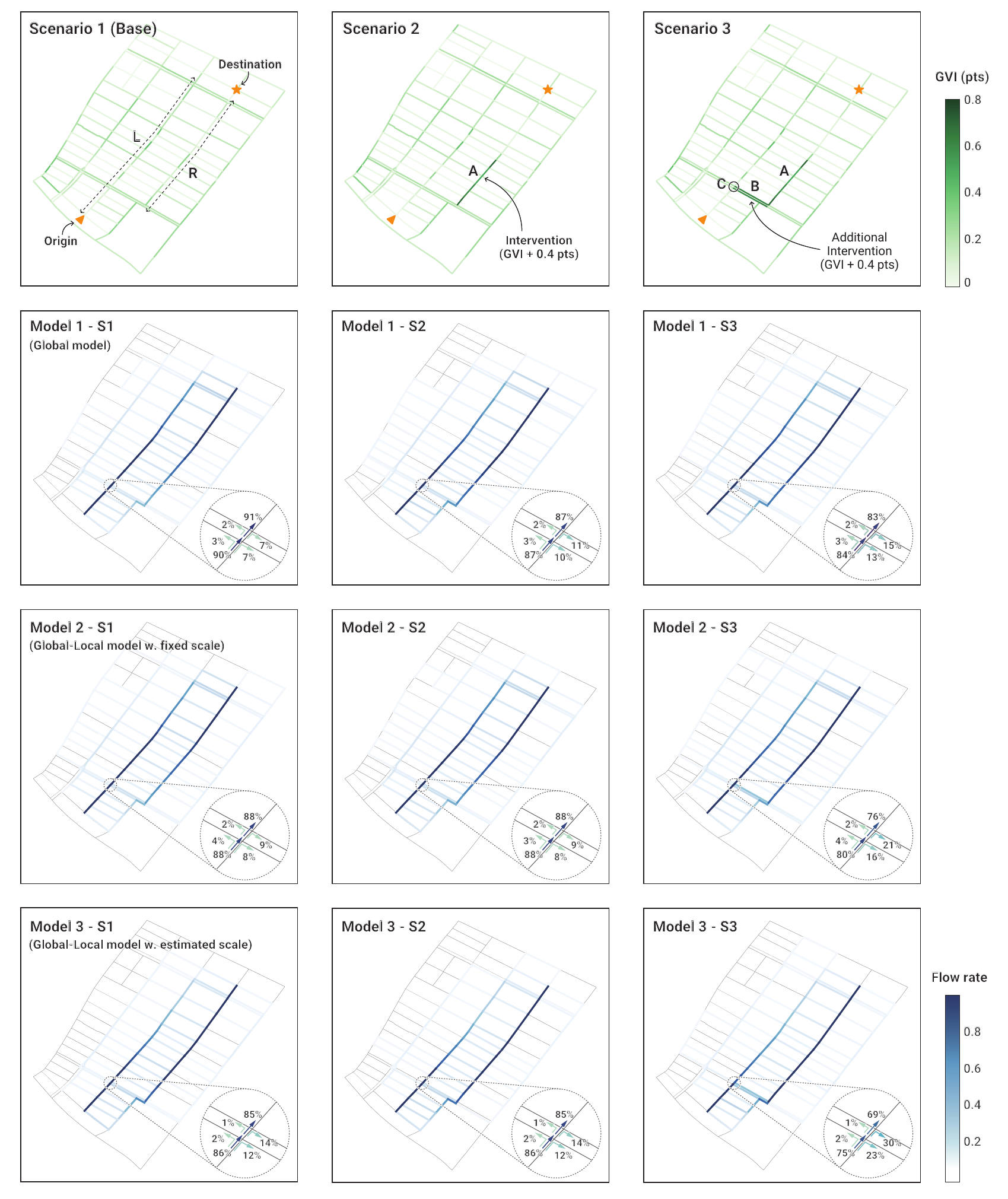}
		\caption{Simulation scenarios and results}
		\label{fig:simulation} 
	\end{center}
\end{figure}

Finally, we present simulation results using the estimated models to analyze how different models predict the change in walking paths according to the increase in GVI. For this analysis, we used a subnetwork of the Kannai network as shown in Figure \ref{fig:simulation}. A single pair of origin and destination, denoted by the triangle and star on the top-left panel, was considered. 
The following three scenarios with different GVI increase policies were evaluated:
\begin{itemize}
    \item \textbf{Scenario 1} (top-left of Figure \ref{fig:simulation}): the base scenario without any intervention 
    \item \textbf{Scenario 2} (top-center of Figure \ref{fig:simulation}): 0.4 pts increase in GVI on Part A 
    \item \textbf{Scenario 3} (top-right of Figure \ref{fig:simulation}): 0.4 pts increase in GVI on both Parts A and B
\end{itemize}
Our main expectation of the policies is to induce pedestrians who walk along Avenue L to Avenue R. 
We performed simulations using the three different models (Models 1-3) in the three different scenarios. The results are shown in the second to fourth rows of Figure \ref{fig:simulation}.

Model 1, the global path choice model, predicted that pedestrians shifted from Avenue L to Avenue R in both Scenarios 2 and 3, indicated by the change in flow rates on Part A. Because Model 1 assumes that pedestrians globally perceive the change in GVI of all streets in the network, the intervention in Part A alone was effective to induce pedestrians to Avenue R. 
In contrast, Models 2 and 3 predicted few changes in flow rates for Scenario 2 compared to Scenario 1. This result suggests that the intervention only on Part A was ineffective to induce pedestrians to Avenue R, because pedestrians perceive the utility associated with the streetscape greenery only locally (visually).

% where we increased the GVI on Part B which is directly connected to Avenue L
In Scenario 3, Models 2 and 3 showed that many pedestrians walked along Avenue R instead of Avenue L. This is because in Scenario 3 we increased the GVI on Part B which is directly connected to Avenue L, and pedestrians walking along Avenue L visually perceived the GVI increase at intersections C and adjusted their path choices. As a result, the link choice probabilities at intersections C changed so that pedestrians are likely to turn right and walk on Part B.
%Part B is directly connected to Avenue L and the GVI increase there was perceived by pedestrians who walk along Avenue L at intersections C. As a result, the local path choice probabilities at intersections C changed so that pedestrians are likely to turn right and pass Part B along their paths.

These results suggest the importance of the location selection of an intervention when travelers perceive the policy variable only locally. 
In such cases, it is effective to introduce a policy on streets/roads connected to parts where people usually travel so that the change is visually perceived and induces their behavioral change.

%%%%%%%%%%%%%%%%%%%%%%%%%%%%%%%%%%%%%
%%%Figure of Simulation Result
%%%%%%%%%%%%%%%%%%%%%%%%%%%%%%%%%%%%%

\section{Conclding remarks}\label{sec:conclusion}
% summary
This study proposed a \textbf{reward decomposition approach} integrated into a link-based recursive logit (RL) path choice model. %for network path choice analysis, which was
%link-based network path choice model with decomposed utilities to analyze both global preferences for and local responses to network attributes. 
The proposed approach decomposes the instantaneous reward function into two utility functions: one is a function of globally perceived attributes, and the other is that of locally perceived attributes. This decomposition allows us to analyze to what extent and which attributes affect the global and local path choice behavior of a traveler.
% global and local utilities, the proposed approach allowed us to analyze to what extent and which attributes affect the global and local preferences of travelers. 
The proposed \textbf{global-local path choice model} can also be estimated from revealed path observations as efficiently as deterministic RL models. The usefulness of the model was demonstrated by the numerical results. %based on synthetic data and real pedestrian path choice observations.
Below we conclude the study by summarizing its main results, potential applications, limitations, and future works.

% In addition to the illustrative examples and a numerical experiment with synthetic data, we applied the proposed model to real pedestrian walking path observations collected in Yokohama-city, Japan. With a semantic segmentation approach, we extracted the GVI from Google StreetView images and used the values as an attribute representing the visual quality of streets. The estimation results showed that the GVI was locally perceived by pedestrians. Moreover, the local path choice models that considered the GVI as a local attribute had better out-of-sample prediction performance than the global path choice model. We also presented several policy implications; in particular, the simulation result suggested the importance of the location selection of interventions when the attribute of interest is locally perceived. 

\subsection{Main results and remarks}
% The usefulness of the model was demonstrated by the numerical results based on synthetic data and real pedestrian path choice observations.
This paper presented two sets of numerical results based on synthetic data and pedestrians' revealed path observations, respectively. 
The experiment with synthetic data in the Sioux Falls network examined the reproducibility of the true parameter values by estimation with different model specifications. The experiment showed that different specifications led to biased estimates and that the estimation of a model introducing the attribute of interest to both the global and local utilities allowed us to analyze which attributes potentially affect the local responses of travelers. 

In the application to the real pedestrian path choice data in Yokohama-city, Japan, the estimation results suggested that pedestrians locally perceive and react to the streetscape greenery (i.e., GVI values), rather than that they have the pre-trip global perception of the GVI values. Moreover, the models with the GVI attribute introduced to the local utility showed higher out-of-sample prediction performance than the global path choice model.
Several policy implications were also obtained through the WTW and simulation analysis. Particularly, through the simulation we discussed the importance of the selection of where the streetscape greenery is increased or newly introduced, and the result suggested that the intervention should be placed on streets that are directly connected to the streets pedestrians often walk on.

In the case study, we focused our interest on the attribute of GVI and compared the different specifications based on it. Yet, the proposed framework allows the analyst to introduce any attributes into both global and local utility functions. As we showed with the example of the GVI attribute, it is possible to analyze to what extent and which attributes affect the local responses of travelers, by estimating and comparing different specifications with respect to the attributes of interest.

While the discount factor $\gamma$ was fixed to one in the presented case studies, it is also possible to integrate a discounted case with the proposed framework. A discounted model describes the trade-off between the current and future utilities, and a significant discount represents the myopic decision-making of a traveler \citep{Oyama2017GRL}. This is a related but different mechanism to the local path choice behavior. The present model assumes that a traveler visually perceives some attributes en route and locally reacts to them, while s/he still globally perceives some other attributes to choose a path efficiently leading to the destination. 
\ref{app:discount} reports the estimation results of the discounted cases with $\gamma < 1$ in the pedestrian path choice application. The result shows that the discounting significantly deteriorated the goodness-of-fit of the models, indicating that pedestrians place importance on their global preferences for the other attributes while locally reacting to the GVI values.

\subsection{Limitations and future works}
While we consider some attributes being locally perceived by travelers, this study deals with a static and deterministic network. 
Because unexpected conditions in a network, including visual street qualities, are often time-dependent, the empirical analysis of travelers' local responses to dynamic attributes would be an important future work. 
This will require a new survey method that combines computer vision technologies to simultaneously collect the data of a path traveled and the environment so that the dynamic network attributes correspond to the time at which a path is observed.
% Collecting the data of various dynamic network attributes that correspond to the time when a path is observed would require a new survey method such as eye-tracking.

Another limitation of the present framework is that it  only captures the linear effect of an attribute. Given that link-based recursive path choice models are mathematically related to the IRL models \citep{ziebart2008maximum, zhao2023deep}, future work could integrate our reward decomposition approach into the deep IRL framework to capture non-linear effects.

% Future research directions
% traffic dynamics \citep{como2013stability}.
% Comparisons to stochastic MDP models \citep{mai2021route}.
% Integrate it into the RL/IRL framework \citep{ziebart2008maximum}.

\subsection{Other potential applications}
The proposed approach succeeded in capturing the local responses of pedestrians to the visual streetscape greenery in the case study. Yet, it is potentially useful for the analysis of many other types of networks in which agents may visually perceive and locally adapt to network conditions, including a disrupted network with unexpected events and a capacitated multi-modal network with shared mobility, as well as non-transportation agents like animals \citep{hirakawa2018can, kivimaki2020maximum}. 
% travelers including bikes and scooters, as well as non-transportation agents like robots and animals who should/may react to the local environment perceived during the movement.
% This paper presented an application to pedestrian path choices as a case study, the proposed framework can be useful for many other types of travelers including bikes and scooters, as well as non-transportation agents like robots and animals, potentially reacting to the local environment during travel. 

Traffic dynamics simulation would be also a potential application as investigated in \cite{como2013stability} and \cite{hoogendoorn2015continuum}. For a multi-agent system, it is often unrealistic and also computationally expensive to consider every agent globally anticipating the future movements of other agents. The global-local path choice model allows us to introduce only static attributes into the global utility and consider the dynamic attributes or interactions in the local utility so that we can avoid the evaluation of the value function many times and for many agents. This would significantly reduce the computational effort of the simulation.

\section*{Acknowledgements}
This work was financially supported by JSPS KAKENHI Grant numbers 20K14899 and 23H01586. The data for the case study was collected through a Probe Person survey, a complementary survey of the Sixth Tokyo Metropolitan Region Person Trip Survey.

% \clearpage
\appendix

\renewcommand{\thelemm}{\Alph{section}\arabic{lemm}}

\section{Derivation of the gradients}\label{app:gradient}
This appendix provides the derivations of the gradients of the log likelihood function (\ref{eq:LL}) with respect to the parameters to be estimated.
The gradient with respect to a specific parameter $\theta$ is:
\begin{align}
    \label{eq:grad_theta}
    \pdv{LL}{\theta} &= \sum^N_{n=1}\sum^{J_n-1}_{j=1} \left\{ 
    \pdv{v(a_{j+1}|a_j)}{\theta} + \pdv{V(a_{j+1})}{\theta} - \sum_{a \in A(a_j)} P(a|a_j) \left( \pdv{v(a|a_j)}{\theta} + \pdv{V(a)}{\theta} \right)
    \right\}
\end{align}
where
\begin{align}
    \pdv{v(a|k)}{\theta} = 
    \left\{
    \begin{array}{c l}
    x^{L,i}_{a|k}, & {\rm if}~~ \theta = \beta^L_{i}\\
    x^{G,h}_{a|k}, & {\rm if}~~ \theta = \beta^G_{h}\\
    0, & {\rm if}~~ \theta = \mu_G\\
    \end{array}
    \right.
\end{align}
% \begin{align}
%     \pdv{V(a)}{\theta} = 
%     \left\{
%     \begin{array}{c l}
%     0, & {\rm if}~~ \theta = \beta^L_{i}\\
%     \pdv{V(a)}{\beta^G_{h}}, & {\rm if}~~ \theta = \beta^G_{h}\\
%     \pdv{V(a)}{\mu_G}, & {\rm if}~~ \theta = \mu_G\\
%     \end{array}
%     \right.
% \end{align}
and $\pdv{V(a)}{\theta} = 0$ if $\theta = \beta^L_{i}$, resulting in (\ref{eq:derivative_spec}) presented in Section \ref{sec:estimation}:
\begin{align}
    \pdv{LL}{\beta^{L}_i} &= \sum^N_{n=1}\sum^{J_n-1}_{j=1} \left\{ x^{L,i}_{a_{j+1}|a_j} - \mathbb{E}_{\mathbf{p}} [\mathbf{x}^{L}_i ~|~ a_j]  \right\} \nonumber\\ 
    \pdv{LL}{\beta^G_{h}} &= \sum^N_{n=1}\sum^{J_n-1}_{j=1} \left\{ x^{G,h}_{a_{j+1}|a_j} + \pdv{V(a_{j+1})}{\beta^G_h} - \mathbb{E}_{ \mathbf{p}} \left[\mathbf{x}^{G}_{h} + \pdv{\mathbf{V}}{\beta^G_h} ~\middle|~ a_j \right]  \right\} \nonumber\\ 
    \pdv{LL}{\mu_{G}} &= \sum^N_{n=1}\sum^{J_n-1}_{j=1} \left\{ \pdv{V(a_{j+1})}{\mu_G} - \mathbb{E}_{\mathbf{p}} \left[\pdv{\mathbf{V}}{\mu_G} ~\middle|~ a_j\right]  \right\}. \nonumber
\end{align}
Moreover, the gradient of the global value function $V(k)$ with respect to $\beta^G_h$ is
\begin{align}
    \pdv{V(k)}{\beta^G_{h}} &= \pdv{}{\beta^G_{h}}
    \left(
    \frac{1}{\mu_G} \ln \sum_{a \in A(k)} \exp\left\{\mu_G (v_G(a|k) + V(a))\right\}
    \right) \nonumber\\
    &= \sum_{a \in A(k)} P_G(a|k) \left( x^{G,h}_{a|k} + \pdv{V(a)}{\beta^G_{h}} \right) \nonumber\\
    &= D^G_h(k) +  \sum_{a \in A(k)} P_G(a|k) \pdv{V(a)}{\beta^G_{h}} 
\end{align}
where $P_G(a|k) = M_{ka}z_a/z_k$ and $D^G_{h}(k) = \sum_{a\in A(k)} P_G(a|k) x^{G,h}_{a|k}$ as explained in Section \ref{sec:estimation}. This further reduces to
\begin{align}
    \pdv{\mathbf{V}}{\beta^G_h} = \mathbf{D}^G_h + 
    \mathbf{P}^\top_G \pdv{\mathbf{V}}{\beta^G_h}
    ~~~\Leftrightarrow~~~
    \pdv{\mathbf{V}}{\beta^G_h} = (\mathbf{I} - \mathbf{P}^\top_G)^{-1} \mathbf{D}^G_h
\end{align}
where $(\mathbf{I} - \mathbf{P}^\top_G)$ is invertible \citep{Baillon2008MCA}. Hence, the gradient can be computed by solving the system of linear equations.

The gradient of the global value function $V(k)$ with respect to $\mu_G$ is
\begin{align}
    \label{eq:grad_V_mu}
    \pdv{V(k)}{\mu_G} &= \pdv{}{\mu_G}
    \left(
    \frac{1}{\mu_G} \ln \sum_{a \in A(k)} \exp\left\{\mu_G (v_G(a|k) + V(a))\right\}
    \right) \nonumber\\
    &= -\frac{1}{\mu_G} \left(
        V(k) - \sum_{a \in A(k)} P_G(a|k) (v_G(a|k) + V(a))
    \right) + \sum_{a \in A(k)} P_G(a|k) \pdv{V(a)}{\mu_G}.
\end{align}
We herein focus on the (conjugate) relationship between the value function $V(k)$ and the entropy function $H(k)$ \citep[e.g.,][]{oyama2022markovian}:
\begin{align}
    H_G(k) &= - \sum_{a \in A(K)} P_G(a|k) \ln P_G(a|k) \nonumber\\
    &= - \mu_G \sum_{a \in A(K)} P_G(a|k) (v_G(a|k) + V(a) - V(k)) \nonumber\\
    &= - \mu_G  \left(
        V(k) - \sum_{a \in A(k)} P_G(a|k) (v_G(a|k) + V(a))
    \right).  
\end{align}
As a result, (\ref{eq:grad_V_mu}) reduces to
\begin{align}
    \pdv{\mathbf{V}}{\mu_G} = - \frac{1}{\mu_G^2} \mathbf{H}_G + \mathbf{P}^\top_G \pdv{\mathbf{V}}{\mu_G}
    ~~~\Leftrightarrow~~~
    \pdv{\mathbf{V}}{\mu_G} = - \frac{1}{\mu_G^2} (\mathbf{I} - \mathbf{P}^\top_G)^{-1} \mathbf{H}_G.
\end{align}

\section{Discounting of global value function}\label{app:discount}
This appendix presents the estimation results of the proposed path choice model with the discounting of the global utility.
The discount factor $\gamma$ represents the trade-off between the current and future utilities of a traveler \citep{Oyama2017GRL}. We fixed $\gamma$ to one, i.e., analyzed only the undiscounted case, for the application to pedestrian path choices in Section \ref{sec:application}, but here we additionally estimated the three models (Models 1--3) with varying $\gamma$ values smaller than one.

\begin{figure}[htb]
	\begin{center}
        \includegraphics[width=15cm]{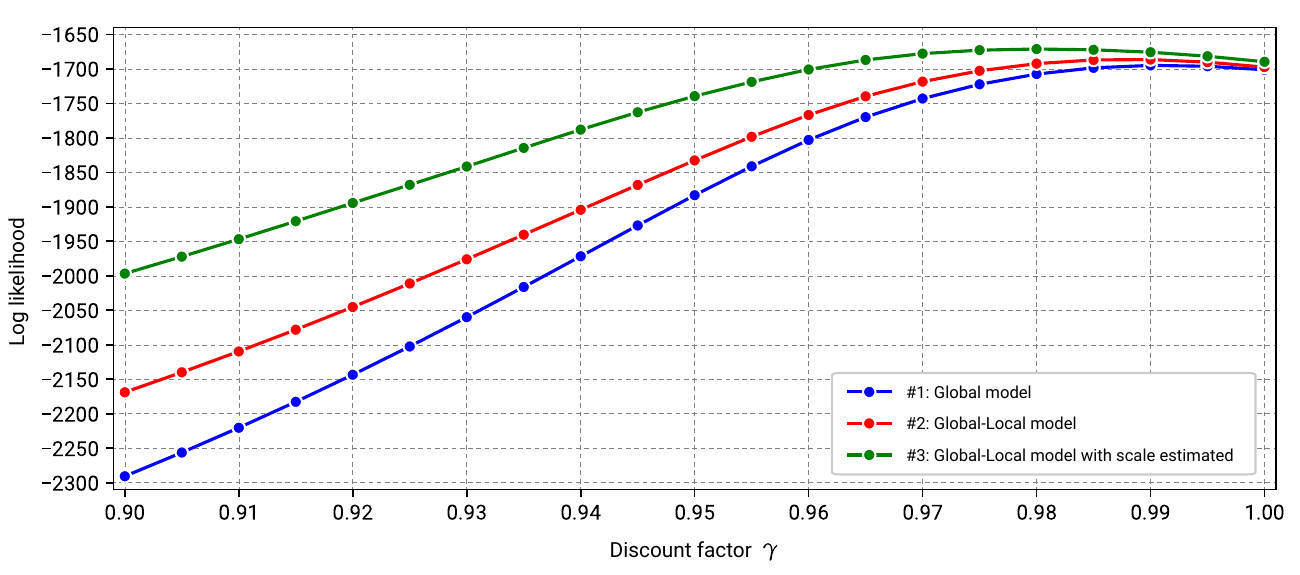}
		\caption{Final log-likelihood values of the discounted models with different values of the discount factor.}
		\label{fig:discount} 
	\end{center}
\end{figure}

Figure \ref{fig:discount} shows the final log-likelihood of the three models for varying discount factor values from 0.90 to 1. The log-likelihood values slightly got better with $\gamma$ between 0.96 and 0.99 than the undiscounted model. However, the models fitted worse with $\gamma$ smaller than 0.96, and the log-likelihood values monotonically decreased as $\gamma$ became smaller. 

These results clearly show that although pedestrians locally perceive some attributes and myopically react to the local environment, it does not mean that they make light of the future in decision-making. In other words, while pedestrians still have global path preferences such as for paths with short lengths or wide sidewalks, at the same time they respond to the visually perceived attributes. 
That is why the discounting of the global utility did not better describe the pedestrians' path choice behavior, and rather, the differentiation of global and local utilities at the attribute level improved the understanding of path choice preferences.

%%%%%%%%%%%%%%%%%%%%%%%%%%%%%%%%%%%%%
%%%Figure of Estimation results
%%%%%%%%%%%%%%%%%%%%%%%%%%%%%%%%%%%%%
%%%%%%%%%%%%%%%%%%%%%%%%%%%%%%%%%%%%%

\bibliography{globalocal}
\bibliographystyle{elsarticle-harv}

\end{document}